\documentclass[11pt,a4paper]{article}
\pdfoutput=1
\usepackage{jcappub}
\usepackage{latexsym}
\usepackage[latin1]{inputenc}
\usepackage{dcolumn}
\usepackage{amsmath,amssymb}        % amssymb includes amsfonts
\usepackage{graphicx}% Include figure files
\usepackage{bm}% bold math
\usepackage{epsfig}

\title{Structure formation with scalar field dark matter: the field approach}

\author{Juan Maga\~na$^{a,1}$\note{Part of the Instituto Avanzado de Cosmolog\'ia (IAC) collaboration http://www.iac.edu.mx/},
Tonatiuh Matos$^{b,1}$Abril Su\'arez$^{b,1}$, F.J. S\'anchez-Salcedo$^{a,1}$}

\affiliation[a]{Instituto de Astronom\'{\i}a,
              Universidad Nacional Aut\'onoma de M\'{e}xico,
              AP 70-264, 04510 M\'exico D.F., M\'exico}

\affiliation[b]{Departamento de F\'isica, Centro de Investigaci\'on y de
  Estudios Avanzados del IPN, A.P. 14-740, 07000 M\'exico D.F.,
  M\'exico.}

\emailAdd{aldebaran.99@gmail.com}
\emailAdd{tmatos@fis.cinvestav.mx}
\emailAdd{asuarez@fis.cinvestav.mx}
\emailAdd{jsanchez@astro.unam.mx}
\abstract{
We study the formation of structure in the Universe assuming that dark matter can be described by a scalar field $\tilde\Phi$ with a potential 
$V(\Phi)=-\mathfrak{m}^{2}\tilde\Phi^{2}/2+\lambda\tilde\Phi^4/4$. 
We derive the evolution equations of the scalar field in the linear regime of perturbations. 
We investigate the symmetry breaking and possibly a phase transition of this scalar field in the early Universe. At low temperatures, 
the scalar perturbations have an oscillating growing mode and therefore, this kind of dark matter could lead to the formation of gravitational structures. 
In order to study the nonlinear regime, we use the spherical collapse model and show that, in the quadratic potential limit, 
this kind of dark matter can form virialized structures. 
The main difference with the traditional Cold Dark Matter paradigm is that the formation of structure in the scalar field model can occur at earlier times. 
Thus, if the dark matter behaves as a scalar field, large galaxies are 
expected to be formed already at high redshifts. 
}


\begin{document}
\maketitle
\flushbottom
\section{Introduction}
%----------------CDM model-------------------------------
One of the main achievements of the Big Bang model is the way it can describe the beginning of our expanding Universe. 
The well known standard model of Cosmology,  $\Lambda$-Cold Dark Matter ($\Lambda$CDM), 
which is based on the Big Bang theory, predicts that $23\%$ of the total
content of the Universe is dark matter. Even though observations of structure formation in the Universe, i. e., galaxies, clusters of galaxies and large-scale structure, give support to this inference, the nature of dark matter 
(DM) remains unknown.

Despite all its successful achievements, there are some aspects of the 
$\Lambda$CDM model which require further consideration.
Shaun et al., \cite{Shaun}, 
found small anomalies in the mass power spectrum (MPS) obtained by SDSS and the one obtained with the $\Lambda$CDM model, 
for small $l$, i. e., anomalies in the predicted large-scale structure of the Universe.  The $\Lambda$CDM 
paradigm faces several challenges to explain observations at 
galactic scales, such as the central densities 
of dark halos in Low Surface Brightness (LSB) galaxies and the excess 
of satellite galaxies predicted by N-body simulations.
In other words, there is not a match between $\Lambda$CDM predictions
at galactic scales and what is being observed. 
Problems with an otherwise successful model are often the 
key to a new and deeper understanding. 

Given these discrepancies, it seems necessary to explore alternatives 
to the paradigm of structure formation. 
Recently, several alternative models have been proposed.
%------------------------SFDM model ---------------------------------
One of them invokes a scalar field as the dark matter (SFDM) in the
Universe \cite{phi2,matos_open}. 
This model supposes that dark matter is a real scalar field (SF) $\Phi$ minimally coupled to gravity that is endowed with a scalar potential $V(\Phi)$, 
and that at some temperature only interacts gravitationally with the rest of the matter. 
It is known that an exponential-like scalar field potential fits very well the cosmological constraints due to the form of its solutions 
(see for example \citep{Wetterich,Ratra,Copeland}). The most simple model 
having both an 
exponential behaviour and a minimum is a $cosh$-like potential. 
\citet{matosurena00b} used a potential of the form $V(\Phi)=V_{0}\left[\cosh\left(\xi\Phi\right) -1\right]$, where $V_{0}$ and $\xi$ 
are constants, to perform a first cosmological analysis in the context of SFDM. 
They showed that the evolution of the Universe, its expansion rate
and the growth of linear perturbations in this model are identical 
as those derived in the standard model. A couple of years ago, 
we developed in \cite{phi2} a formalism to show that a scalar field 
with a quadratic potential $V(\Phi)=m^{2} \Phi^{2}/2$ can reproduce the cosmological evolution of the Universe. 
An interesting result is that the predicted density of neutrinos at the recombination epoch is in agreement 
with the observations of the Wilkinson Microwave Anisotropy Probe (WMAP).
Recently Su\'arez et al.~\cite{suarez} developed a hydrodynamical approach for the structure formation in the Universe 
with the scalar potential $V(\Phi)=m^{2}\Phi^{2}/2 + \lambda\Phi^{4}/4$. 
They found that when $\lambda=0$ the evolution of the perturbations of the SFDM model compared to those of $\Lambda$CDM are nearly identical.
They also showed 
that this potential can lead to the early formation of gravitational structures in the Universe if the self-interaction parameter
$\lambda$ is $<0$.

%---------------------------BEC/SFDM----------------------------------------------------
It has also been proposed that this dark matter scalar field, i.e., this spin-0 fundamental interaction, could lead to 
the formation of Bose-Einstein condensates (BECs) in the way of cosmic structure \cite{matosurena00b,dehnen,harko_cosmo,harkomnras11,chavanisIII}. 
In the non self-interacting case, SFDM forms BECs if the mass of the associated particle, $m$, is $<10^{-17}$eV \cite{matosurena00b, hu}.
The thermodynamical analysis of BEC indicates that gravitational structures of SFDM can be 
formed at earlier times than CDM structures \cite{phi2}. In a recent paper, \citet{luis_bose} studied the conditions 
for the formation of a SFDM/BEC in the Universe, also concluding that SFDM/BEC particles must be ultra light bosons.
In the same direction, \citet{ivan} studied ultra light bosons as dark matter in the Universe
with the framework of kinetic theory, through the Boltzmann-Einstein equations, and they found that this kind
of ultra light particles is consistent with the acoustic peaks of the cosmic microwave background radiation
if the boson mass is $m\sim10^{-22}$eV.

%---------------------general numerical results-------------------------------------------
\cite{leelimchoi} pointed out that SFDM/BEC can explain the 
spatial separation of the dark matter from visible matter,
as derived from 
X-ray maps and weak gravitational lensing, 
in the Bullet Cluster \cite{clowe}.
On the other hand, several authors have numerically studied the formation, collapse and virialization of SFDM/BEC halos as well as
the dynamics of the SFDM around black holes \citep{fcoluis,colapsob,apjpaco,woo,chavanisII,gonzalez,cruz-osorio,barranco}. 
\citet{alcubierre} found that the critical mass for collapse is of the order of a Milky Way-sized halo mass.
This suggests that SFDM/BEC can be a plausible candidate 
to dark matter in galactic halos.  
%---------------------galactic scales-------------------------------------------
Other studies show that SFDM/BEC predicts intriguing phenomena at galactic scales. 
\citet{victor} (see also \citep{boharko,argelia,harko_core})
showed that BEC dark matter halos fit very well high-resolution rotation curves of LSB galaxies, and how the constant density core in dark halos
can be reproduced.
Also, \cite{leelim} shows that the SFDM/BEC paradigm is a good alternative to explain the common mass of the dark halos of 
dwarf spheroidal galaxies.
Recently, \citet{rindler} investigate the formation of vortex in SFDM/BEC halos. They found constraints
on the boson mass in agreement with the ultra light mass found in previous works (see also \citep{kaina,zinner}).
In addition, \citet{lora} studied, through N-body simulations, the dynamics of Ursa Minor dwarf galaxy and its stellar clump 
assuming a SFDM/BEC halo to establish constraints for the boson mass.
Moreover, they introduced a dynamical friction analysis with the SFDM/BEC model 
to study the wide distribution of globular clusters in Fornax. An overall good agreement is found
for the ultra light mass $\sim 10^{-22}$eV of bosonic dark matter.

In this work we consider the hypothesis that dark matter
can be described by self-interacting scalar field under the action of a potential that goes as 
$V(\tilde\Phi)=-\mathfrak{m}^2\tilde\Phi^2/2 +\lambda\tilde\Phi^4/4$, 
and is affected by gravity in an indirect way through the gravitational potential $\phi$. 
The journey of this SF starts at very early stage of the Universe, where the temperature effects are very relevant for
the scalar dynamics. At this high temperature regime, the SF has a symmetry breaking and then it reaches one of its minima and its scalar
potential becomes quadratic. When the temperature decreases by the expansion of the Universe, the SF becomes the main dark matter component in the Universe 
evolving in the same way as CDM. Therefore, we are interested in the growth of the scalar perturbations in both linear and nonlinear regimes
to investigate the main differences with those of the standard paradigm.

This paper is organized as follows. In section \ref{sec:pot}, we give some properties on the behavior of the $\lambda\tilde\Phi^4$ potential with temperature contributions. 
In section \ref{sec:background}, we study the behavior of the background, when $\Phi$ and other cosmological parameters depend only on time. 
In \ref{sec:linear} we analyse the linear behavior of the perturbations, 
while  the nonlinear regime is discussed
in section \ref{sec:qlinear}. 
The majority of these sections are developed when the SF has reached one of its minima, adopts a $\Phi^2$ potential profile and the 
mass term becomes positive at the end of the phase transition. Finally,
conclusions are given in section \ref{sec:conclusions}.

\section{The Cosmological Scalar Field Potential}
\label{sec:pot}

To study the cosmological dynamics of the SFDM model we consider the simplest case: a single scalar field $\tilde\Phi(x,t)$, 
with self-interacting double-well potential (Mexican-hat potential). We write the potential as 
\begin{equation}
V(\tilde\Phi)= \frac{\lambda}{4}\left(\tilde\Phi^{2} - \frac{\mathfrak{m}^2}{\lambda}\right)^2.
\label{mexicanhat_sT}
\end{equation}
In a very early stage of the Universe, 
this scalar field was in local thermodynamic equilibrium with its surroundings, see \cite{kolb,Briscese:2011ka,Briscese:2011zz}.  
At some time, the scalar field decoupled from the rest of the matter and started a 
lonely journey with its temperature $T$ going down by the expansion of the Universe. Thus, we consider the scalar field in a thermal bath of temperature $T$,
whose scalar field potential, extended to one loop corrections, is given by
(in natural units $c=\hbar=k_{B}=1$)
\begin{equation}
V(\tilde\Phi)= -\frac{1}{2}\mathfrak{m}^{2}\tilde\Phi^{2} + \frac{\lambda}{4}\tilde\Phi^{4}+\frac{\lambda}{8}T^2\tilde\Phi^2-\frac{\pi}{90}T^4+\frac{\mathfrak{m}^4}{4\lambda},
\label{mexicanhat}
\end{equation}
where $\mathfrak{m}$ is a mass parameter and $\lambda$ is the self-interacting constant. We can calculate the critical temperature $T_{c}$ at which the $Z_2$ symmetry of our real SF breaks. In order to do that, we obtain the critical points of the scalar potential (\ref{mexicanhat}) as

$$0=(-\mathfrak{m}^2+\lambda\tilde\Phi^2+\frac{\lambda}{4}T^2)\tilde\Phi$$.
It is important to note that the negative term $-\mathfrak{m}^{2}$ permits the breaking of symmetry of our potential. The first critical point is found at $\tilde{\Phi}=0$. If the temperature $T$ is high enough, the scalar potential (\ref{mexicanhat}) has a minimum at this critical point. Furthermore, the critical temperature $T_{c}$ at which $V$ has a maximum at $\tilde{\Phi}=0$ is 
$$T_c^2=\frac{4\mathfrak{m}^2}{\lambda}.$$ This critical temperature defines the symmetry breaking scale of the scalar field. 

As the temperature drops, the minimum of $V$ occurs when $\tilde\Phi\neq 0$,
$$\tilde\Phi=\pm\frac{1}{2}(T_c^2-T^2)^{1/2}.$$

As the scalar field passes through $T_c$, there are local thermal fluctuations of the field that will drive it from the unstable maximum of 
$V$ at $\tilde\Phi=0$ towards one or other of the minima. After the scalar field passes through the breaking of symmetry 
(and possibly a phase transition), the scalar potential is stabilized and begins to oscillate around its minimum.  

Let us choose the positive minima, $\Phi_{min+}$, of the scalar field (remember there are two minima in the potential) 
and write $\tilde\Phi=\Phi_{min+}+\Phi$. Expanding the potential in $\Phi$, we find after some algebra that
\begin{eqnarray}
V(\Phi_{min+}+\Phi)&=&\left[-\frac{1}{2}\mathfrak{m}^2T_c\left(1-\frac{T^2}
{T_c^2}\right) ^{1/2}
\frac{\lambda}{8}T_c^3\left(1-\frac{T^2}{T_c^2}\right)^{3/2}+\frac{\lambda}{8} T^2T_c\left(1-\frac{T^2}{T_c^2}\right)^{1/2}\right]\Phi\nonumber\\
&+&\left[-\frac{1}{2}\mathfrak{m}^2+\frac{3}{8}\lambda T_c^2\left(1-\frac{T^2}{T_c^2}\right)+\frac{\lambda}{8}T^2\right]\Phi^2
+\frac{\lambda}{2}T_c\left(1-\frac{T^2}{T_c^2}\right)^{1/2}\Phi^3+\frac{\lambda}{4}\Phi^4\nonumber\\
&+&\left[-\frac{1}{8}\mathfrak{m}^2T_c^2\left(1-\frac{T^2}{T_c^2}\right)+\frac{\lambda}{64}T_c^4\left(1-\frac{T^2}{T_c^2}\right)^2
\frac{\lambda}{32}T^2T_c^2\left(1-\frac{T^2}{T_c^2}\right)+\frac{\mathfrak{m}^4}{4\lambda}\right].\nonumber\\
&&
\label{potexp}
\end{eqnarray}
At this point $\mathfrak{m}$ becomes a very relevant parameter. To find
the real value of $\mathfrak{m}^2$ at low temperatures, we have to analyse the value of $\mathfrak{m}^2$ in the original potential at temperatures well below the critical temperature, $T\ll T_c$ (with $T\approx 0$), 
as to say that the system has undergone the breaking of symmetry. With these considerations at hand, we have from eq.~(\ref{potexp}) 
\begin{equation}
V(\Phi_{min+}+\Phi)=\frac{1}{2}m^2\Phi^2+\sqrt{\frac{\lambda}{2}}m\Phi^3+\frac{\lambda}{4}\Phi^4,
\label{potmin}
\end{equation}
where we have defined the positive mass $m$ in the minimum of the potential as $m^2=2\mathfrak{m}^2$.
% and it is at this point where we begin our cosmological analysis. 

Once the scalar field reaches the minimum and keeps oscillating around it, we 
assume that the scalar field acquires very 
small values, i.e, $\Phi\sim 0$, but never reaches zero, 
hence the terms that go as $\Phi^{3}$ and $\Phi^{4}$ in our expansion can be neglected. Also, if we want this kind of dark matter 
to form structures of around $10^{12}M_{\odot}$, then we need the scalar field 
mass to be around $m\lesssim 1$ eV, ( \cite{seidel,ivan2}) 
and $\lambda\sim 10^{-8}$ (in general $\lambda$ always takes small values). Again we are left with a $\Phi^2$ type potential, 
but now with the positive mass term. 
If we take these values, then the critical temperature is $T_c\sim 2000$ eV, 
which is low enough for the scalar field to decouple from the rest of the 
matter, but high enough so that we are still far behind the matter dominated era. Therefore, we have two regimes of the 
cosmological scalar potential at different stages of the Universe

\begin{equation}
V =\left\{
\begin{array}{lr}
-\frac{1}{2}\mathfrak{m}^{2}\tilde\Phi^{2} + \frac{\lambda}{4}\tilde\Phi^{4}+\frac{\lambda}{8}T^2\tilde\Phi^2-\frac{\pi}{90}T^4+\frac{\mathfrak{m}^4}{4\lambda}  &T \sim T_{c}\\ 
\frac{1}{2}{m}^{2}\Phi^{2} &T \ll T_{c}
\end{array}
\right. .
\end{equation}
Note that $\tilde\Phi$ is the scalar field near to symmetry breaking scale. In the following, we investigate if the SFDM endowed with this scalar potential at $T\ll T_c$ is able to mimic
the dynamics of the background Universe predicted by the concordance $\Lambda$CDM model.

\section{Background Universe}
\label{sec:background}

\subsection{The field approach}

First, we study the dynamics of the SFDM model in the background Universe. In order to do so, 
we consider the Friedmann-Lema\^itre-Robertson-Walker (FLRW) metric with scale factor $a(t)$. Our background Universe is composed by
SFDM ($\Phi_{0}$) endowed with a scalar potential $V\equiv V(\Phi_0)$, baryons ($b$), radiation ($z$), neutrinos ($\nu$), 
and a cosmological constant ($\Lambda$) as dark energy. We begin by recalling the basic background equations. 
From the energy-momentum tensor $\mathbf{T}$ for a scalar field, the scalar energy density $T_0^0$ and the 
scalar pressure $T_j^i$ are given by
\begin{equation}
T_0^0=-\rho_{\Phi_0}=-\left(\frac{1}{2}\dot{\Phi}_0^2+ V\right),
\label{rhophi0}
\end{equation}
\begin{equation}
T_j^i=P_{\Phi_0}=\left(\frac{1}{2}\dot{\Phi}_0^2-V\right)\delta_j^i,
\label{pphi0}
\end{equation}
where the dots stand for the derivative with respect to the cosmological time and $\delta^i_j$ is the Kronecker delta. Thus, the Equation of State (EoS) for the scalar field is $P_{\Phi_{0}}=\omega_{\Phi_{0}}\,\rho_{\Phi_0}$ with
\begin{equation}
\omega_{\Phi_{0}}= \frac{\frac{1}{2}\dot{\Phi}_{0}^{2}\,-\,V}{\frac{1}{2}\dot{\Phi}_{0}^{2}\,+\,V}.
\label{ec:w}
\end{equation}

The radiation fields, the baryonic component and the cosmological constant are represented by perfect fluids with barotropic equation of state $P_{\gamma}=(\gamma-1)\rho_{\gamma}$, where $\gamma$ is a constant, $0\le \gamma \le 2$. For example, $\gamma_{z}=\gamma_{\nu}=4/3$ 
for radiation and neutrinos, $\gamma_{b}=1$ for baryons, and for a cosmological constant $\gamma_{\Lambda}=0$.

The Einstein-Klein-Gordon equations that describe this Universe are
\begin{subequations}
\begin{eqnarray}
\dot H&=&-\frac{\kappa^2}{2}\left(\dot{\Phi}_{0}^2+\frac{4}{3}\rho_z+\frac{4}{3}\rho_{\nu}+
\rho_{b}\right),\\
\ddot{\Phi}_{0} &+& 3\,H \dot{\Phi}_{0}+ V,_{\Phi_{0}}=0,\label{eq:backsf0}\\
{\dot\rho_{z}}&+&4\,H \rho_{z}=0,\\
{\dot\rho_{\nu}}&+&4\,H \rho_{\nu}=0,\\
{\dot\rho_{b}}&+& 3\,H \rho_{b}=0, 
\end{eqnarray}\label{eq:back}
\end{subequations}
with the Friedmann constraint
\begin{equation}
H^2=\frac{\kappa^2}{3}\left(\rho_{\Phi_{0}}+\rho_{z}+
\rho_{\nu}+\rho_{b} + \rho_{\Lambda} \right),
\label{eq:FC}
\end{equation}
\noindent
being $\kappa^{2} \equiv 8\pi G$, $H \equiv \dot{a}/a$ the Hubble parameter and the commas stand for the derivative with respect to scalar field. Notice that background scalar quantities at zero order have the subscript $0$.

In order to solve the system of equations (\ref{eq:back}), we define the following dimensionless variables
\begin{eqnarray}
x&\equiv& \frac{\kappa}{\sqrt{6}}\frac{\dot{\Phi}_{0}}{H},\qquad
u \equiv \frac{\kappa}{\sqrt{3}}\frac{\sqrt{V}}{H},\qquad
b\equiv \frac{\kappa}{\sqrt{3}}\frac{\sqrt{\rho_{b}}}{H},\nonumber\\
z&\equiv& \frac{\kappa}{\sqrt{3}}\frac{\sqrt{\rho_{z}}}{H},\quad\;\,
\nu \equiv \frac{\kappa}{\sqrt{3}}\frac{\sqrt{\rho_{\nu}}}{H},\quad\;\;\;
l \equiv \frac{\kappa}{\sqrt{3}}\frac{\sqrt{\rho_{\Lambda}}}{H}.
\label{eq:varb}
\end{eqnarray}
Here, we take the scalar potential for $T \ll T_{c}$, as $V=m^{2}\Phi_{0}^{2}/2$, where $m$, the mass 
of the ultra-light boson particle is $\sim1 \times10^{-23}$eV. Using these variables, 
the equations (\ref{eq:back}) for the evolution of the background Universe are transformed into
\begin{subequations}
\begin{eqnarray}
x'&=& -3\,x - s u+\frac{3}{2}\Pi\,x, \label{eq:dsb_x}\\
u'&=& s x +\frac{3}{2}\Pi\,u, \label{eq:dsb_u}\\
b'&=&\frac{3}{2}\left(\Pi -1\right)\,b, \label{eq:dsb_b}\\
z'&=&\frac{3}{2}\left(\Pi-\frac{4}{3} \right)\,z, \label{eq:dsb_z}\\
\nu'&=&\frac{3}{2}\left(\Pi-\frac{4}{3} \right)\,\nu, \label{eq:dsb_nu}\\
l'&=&\frac{3}{2}\Pi\,l, \label{eq:dsb_l}\\
s'&=&\frac{3}{2}\Pi\,s, \label{eq:dsb_s}
\end{eqnarray}\label{eq:dsb}
\end{subequations}
where the prime denotes a derivative with respect to the e-folding number $N=\ln a$, and $\Pi$ is defined as
\begin{equation}
-\frac{\dot H}{H^2}=\frac{3}{2}(2x^2+b^2+\frac{4}{3} z^2+\frac{4}{3} \nu^2) \equiv \frac{3}{2}\Pi. 
\end{equation}
\begin{figure}
 \centering
 \scalebox{0.6}{\includegraphics{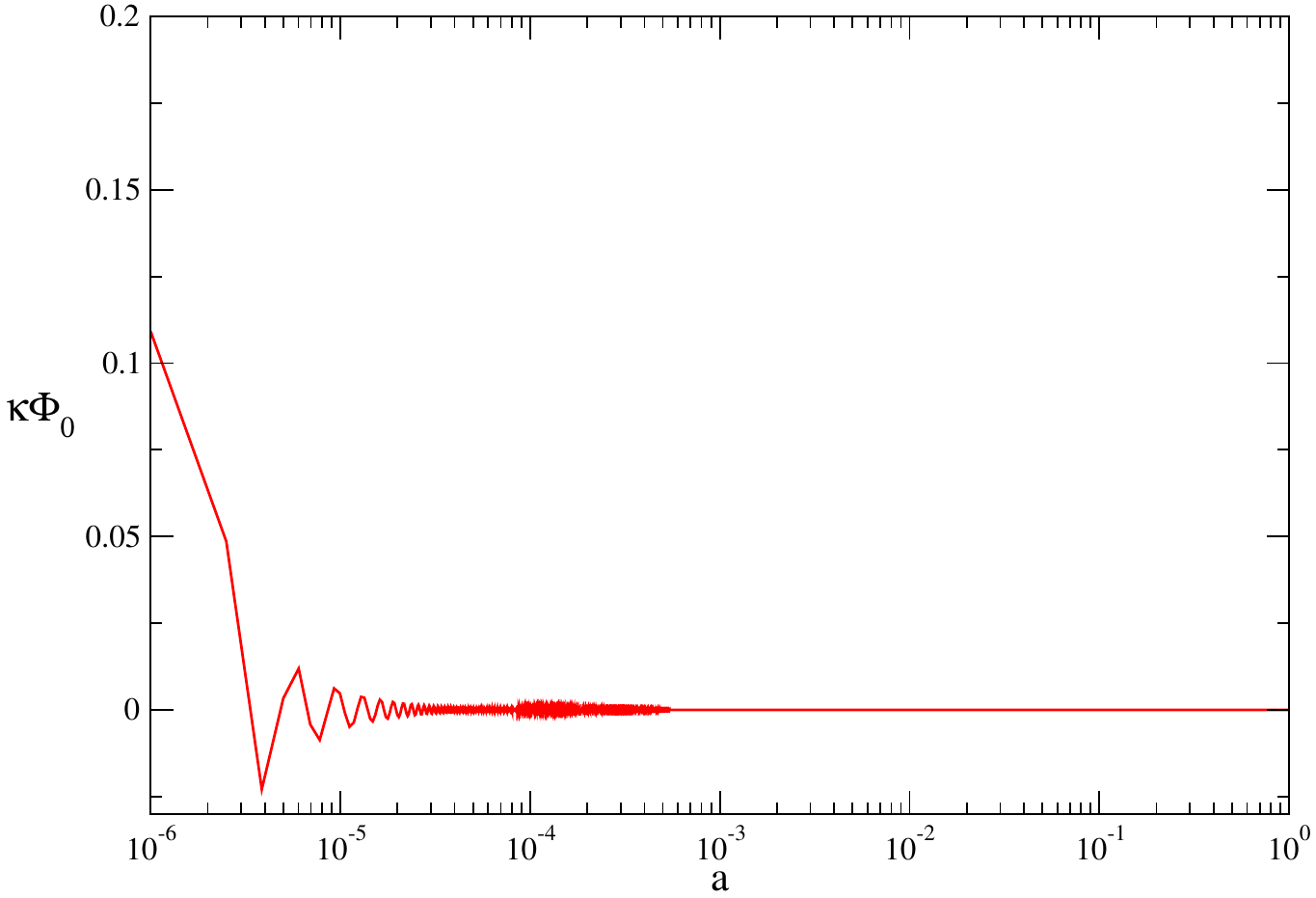}}
\caption{Evolution of the scalar field $\Phi_{0}$ for the background Universe.}
 \label{fig:phi}
\end{figure}
Notice that we have introduced the extra variable $s\equiv m/H$. With these variables, the density parameters $\Omega_{i}$ 
for each component $i$ can be written as
\begin{eqnarray}
\Omega_{\Phi_{0}}&=&x^2+u^2, \qquad \Omega_{b}=b^2, \qquad \Omega_{z}=z^2,\nonumber\\
\Omega_{\nu}&=&{\nu}^2,\qquad\quad\;\;\;\; \Omega_{\Lambda}=l^2,\label{eq:dens}
\end{eqnarray}
subject to the Friedmann constraint
\begin{equation}\label{eq:fri}
x^2+u^2+z^2+\nu^{2}+b^2+l^2=1.
\end{equation}
\begin{figure}
 \centering
 \scalebox{0.6}{\includegraphics{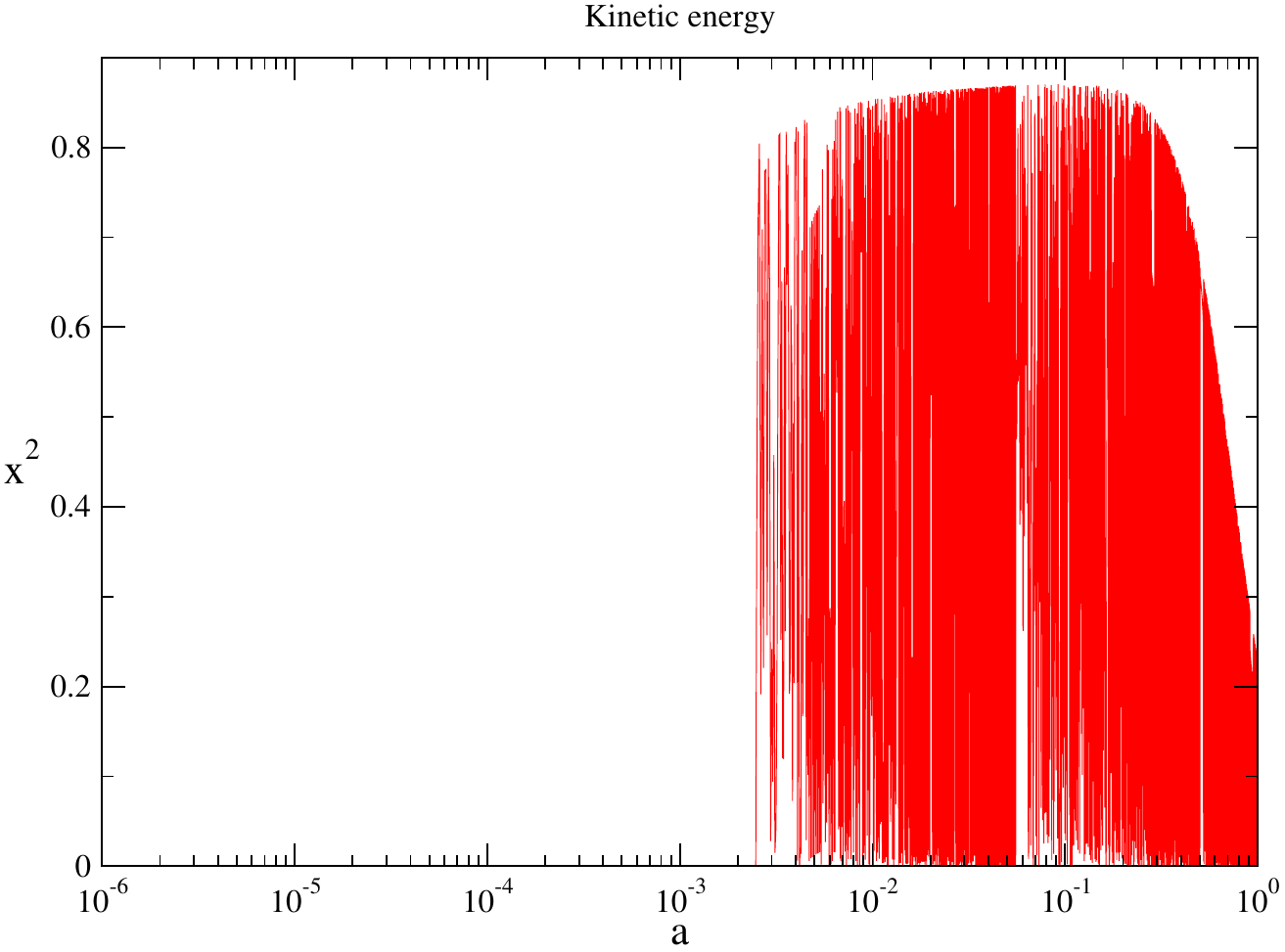}}
  \scalebox{0.6}{\includegraphics{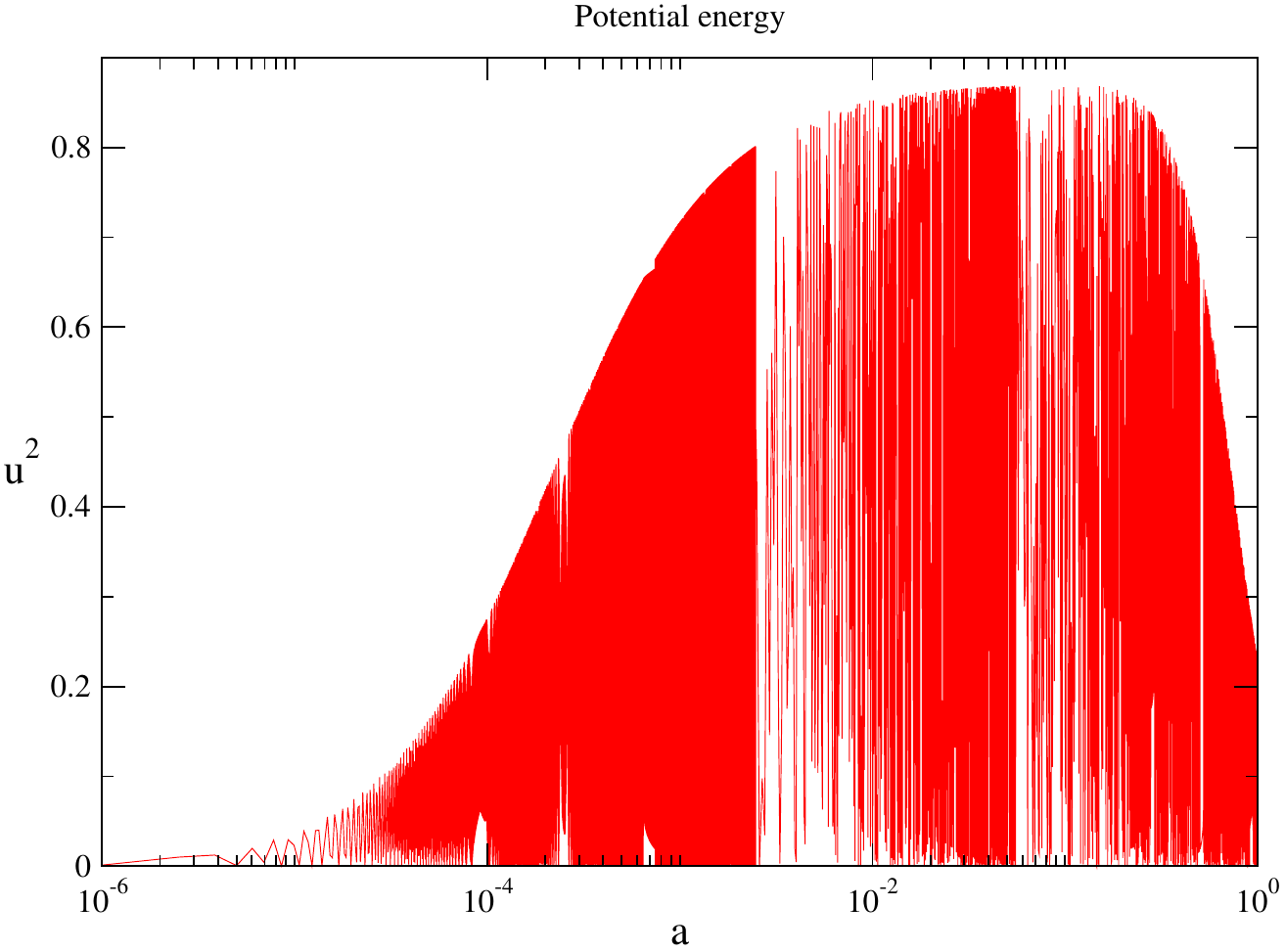}}
 \caption{Evolution of the kinetic (top panel) and potential (bottom panel) energy of the scalar field $\Phi_{0}$.}
 \label{fig:Ox}
\end{figure}
In addition, we may write the EoS of the scalar field as
\begin{equation}
\omega_{\Phi_{0}}=\frac{x^{2}-u^{2}}{\Omega_{\Phi_{0}}}.
\label{eq:dlw}
\end{equation}
$\omega_{\Phi_{0}}$ is a function of time. If the temporal average 
of $\omega_{\Phi_{0}}$ tends to zero, then $\Phi^2$-dark matter can 
be able to mimic the EoS of CDM.

\begin{figure}[ht]
 \centering
 \scalebox{0.6}{\includegraphics{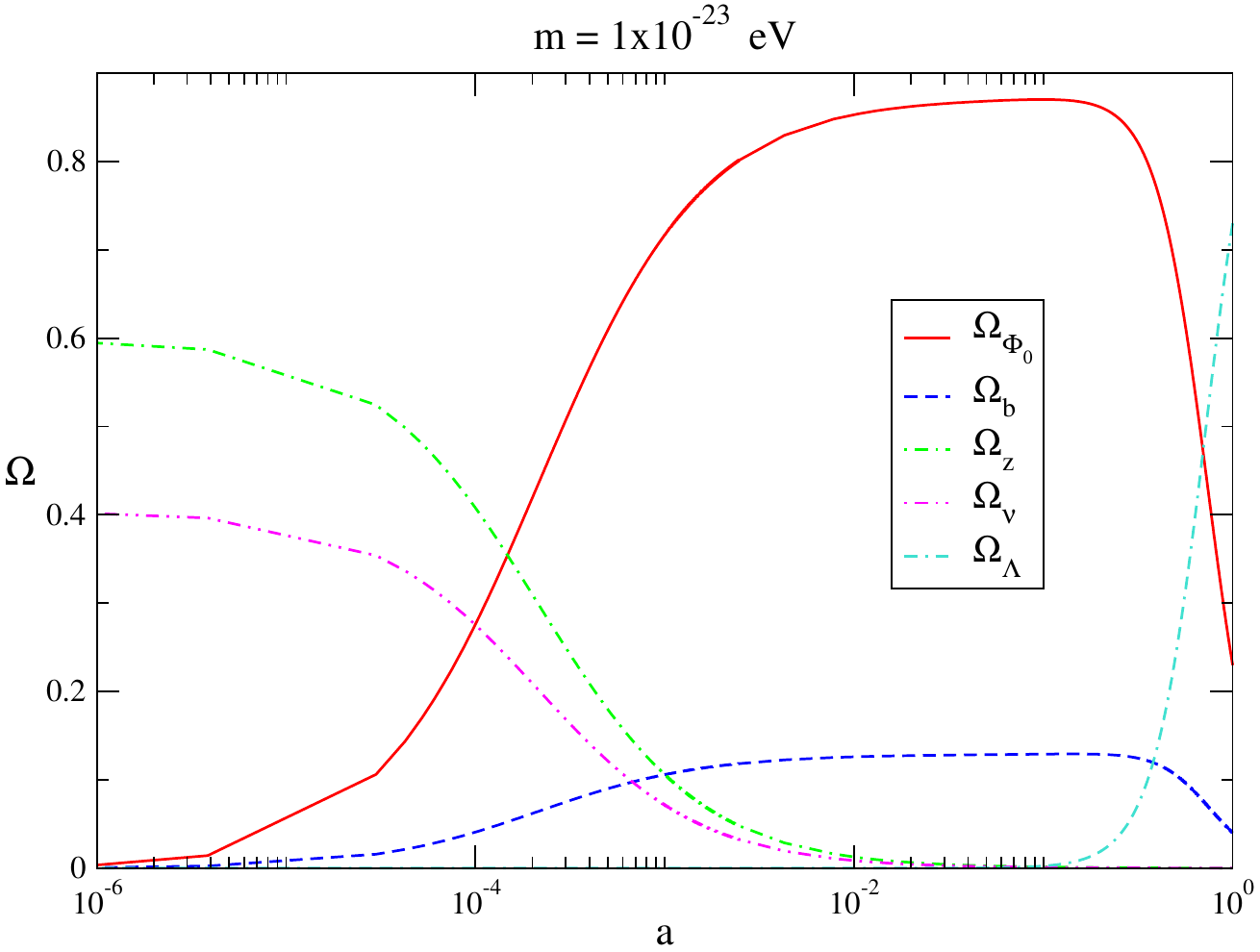}}
 \caption{Evolution of the density parameters $\Omega_{i}$ for the background Universe. Scalar field dark matter model mimics the standard  $\Lambda$CDM behavior.}
 \label{fig:back}
\end{figure}

We solve the system of equations (\ref{eq:dsb}) for the background Universe numerically with an 
appropiate semi-implicit extrapolation method for the resulting stiff system.
We take, as first approach, the initial conditions given by the best estimates from $5$ and $7$-years WMAP values \cite{WMAP5,WMAP7} 
to $\Omega^{(0)}_{\Lambda}=0.73$, $\Omega^{(0)}_{DM}=0.22994$, $\Omega^{(0)}_b=0.04$, $\Omega^{(0)}_z=0.00004$, 
$\Omega^{(0)}_{\nu}=0.00002$, $H_{0}=70$ km s$^{-1}$Mpc$^{-1}$ 
and $m=1\times10^{-23}$ eV that implies $s_{0}=6.65\times10^{9}$. 
In Fig.~\ref{fig:phi} we see the cosmological evolution of the scalar field $\Phi_{0}$. 
This figure shows how the scalar field oscillates about the minimum $\Phi_{0}\sim0$ of the scalar potential $V$. 
In Fig.~\ref{fig:Ox} we show the evolutions of $x^{2}$ and $u^{2}$,
which are related to the 
kinetic ($\dot\Phi_{0}^{2}/2$) and potential energy ($m^2\Phi_{0}^{2}/2$) of the scalar field, respectively.  
As expected, the oscillations are translated into very stark oscillations for the kinetic and the potential energies of 
the scalar field. However, observe the evolution of the dark matter density of the scalar field, 
$\rho_{\Phi_{0}}=(\dot\Phi_{0}^2+m^{2}\Phi_{0}^{2})/2$, in 
Fig.~\ref{fig:back}. A crucial point here is that the kinetic and 
potential energies show very stark oscillations but 
the sum of both energies, that is the density parameter $\Omega_{\Phi_{0}}$,
does not oscillate at all. 
In fact, the oscillations are not physical observables at all; they are a feature of the scalar field. 
What we observe is the density of the scalar field which does not oscillate. 

Fig.~\ref{fig:back} shows the numerical evolution of the density parameters in our model
\footnote{The scale factor is such that $a=1$ today, so that it relates to the redshift $z$ by $a=(1+z)^{-1}$.}. 
At early times, radiation dominates the evolution of the Universe. Later on, the Universe has an epoch where 
the energy density radiation is equal to the dark matter density, at $z_{eq}\sim6000$, 
then dark matter begins to dominate the evolution. The recombination era in scalar field dark matter model occurs at $z\sim1000$. 
At later times, the cosmological constant dominates the dynamics of the Universe at $z_{\Lambda} \sim 0.7$. 
The cosmological behavior of the Universe with SFDM hypothesis is exactly the same as in the $\Lambda$CDM model.

\begin{figure}[ht]
 \centering
 \scalebox{0.6}{\includegraphics{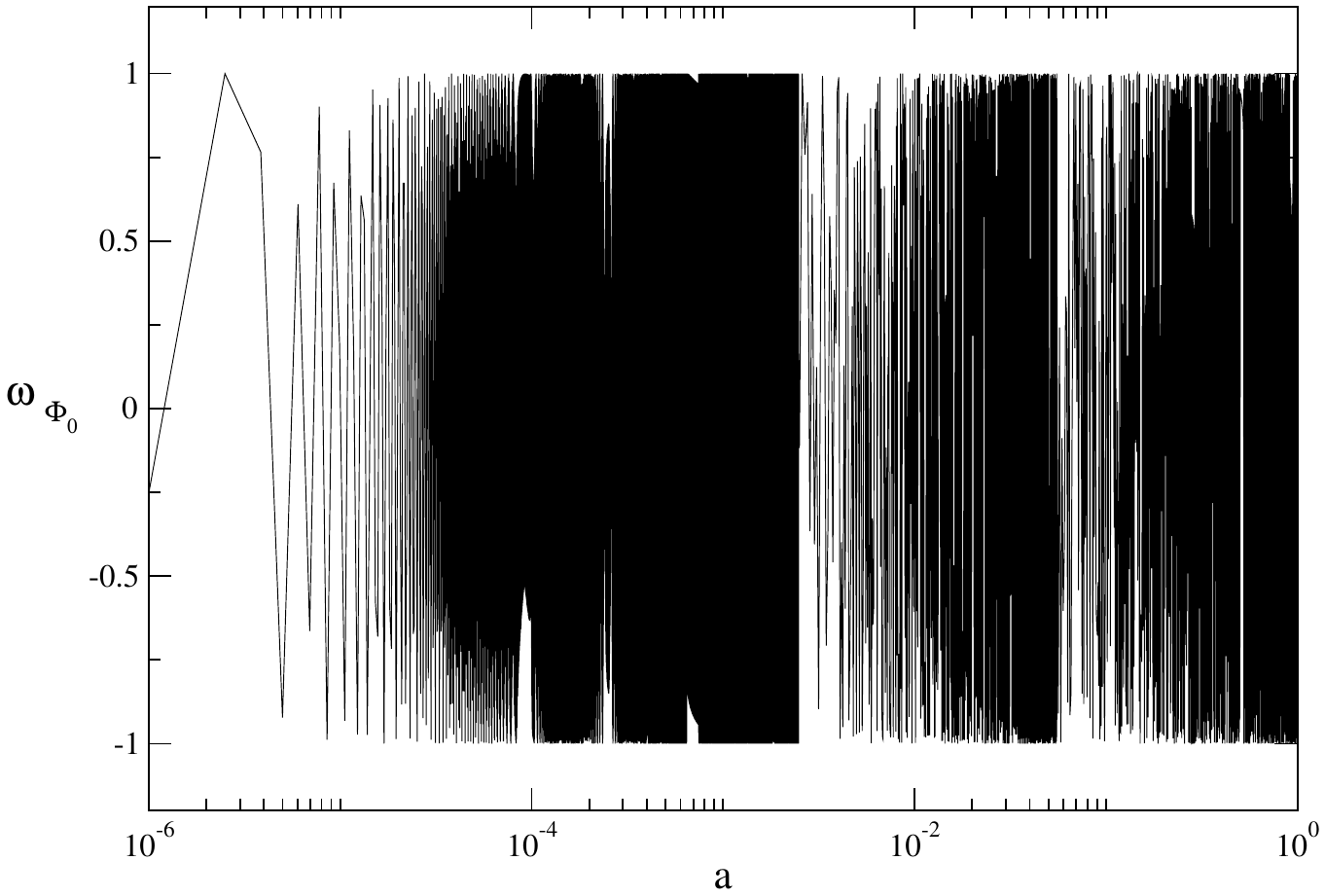}}
 \caption{Evolution of the scalar field dark matter equation of state for the background Universe.}
 \label{fig:w0}
\end{figure}

Fig.~\ref{fig:w0} shows the evolution of the EoS for the scalar field. Although the EoS varies with time (oscillations), 
the temporal average, $\left<\omega_{\Phi_0}\right>$, drops to zero. Therefore, $\Phi^{2}_{0}$ is like a pressureless fluid and 
behaves as cold dark matter at cosmological scales \cite{turner, matosurena00b, phi2, review}.

\subsection{Analytical study of the background Universe}

In this subsection we perform a transformation in order to solve the Friedmann equations with analytic methods with the 
approximation $m \gg H$ and compare this result with the numerical ones. Here the scalar field and the variables of the background depend only on time, 
e.g., $\Phi=\Phi_0(t)$. 

\begin{figure}[ht]
 \centering
 \scalebox{0.6}{\includegraphics{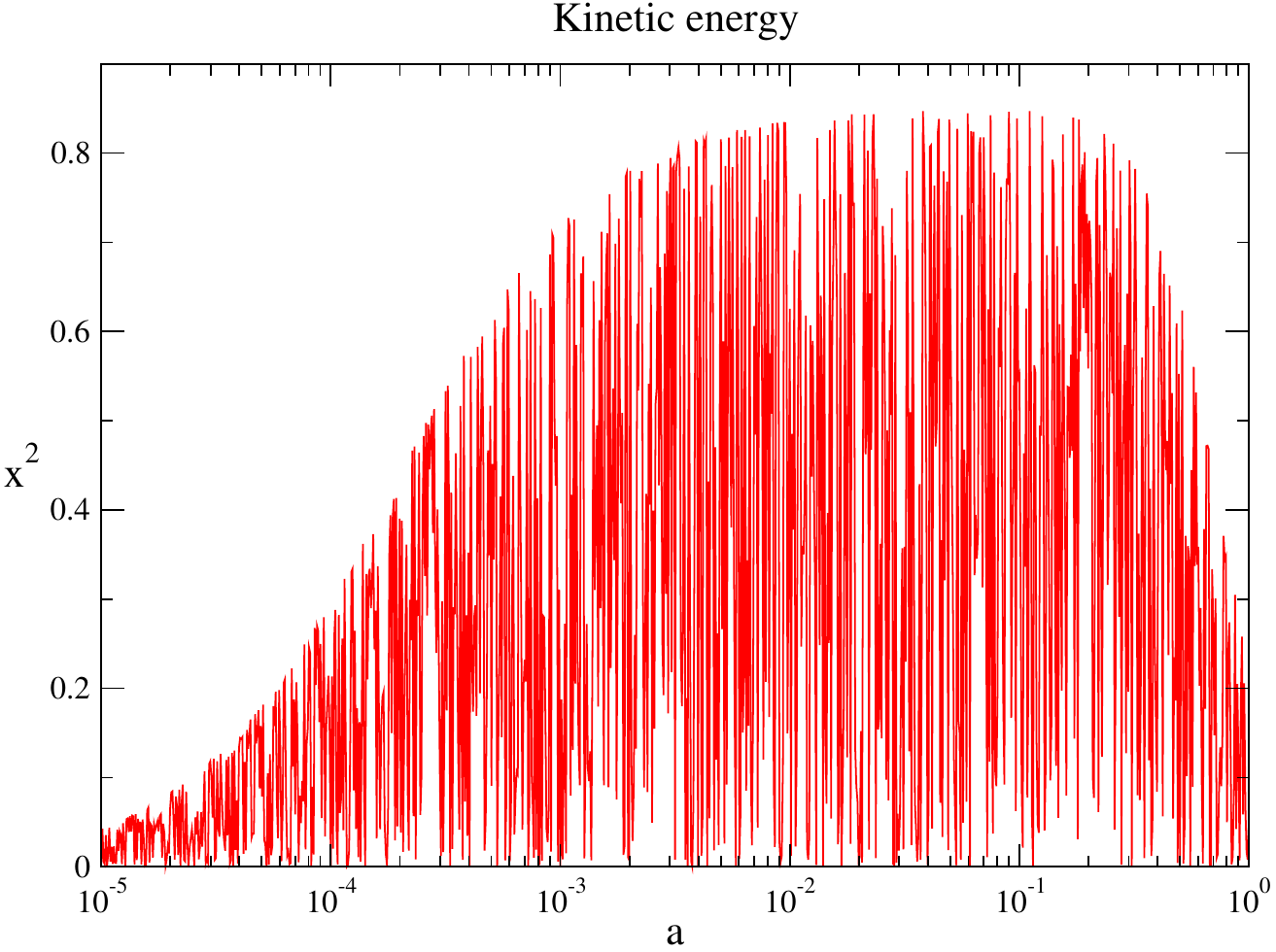}} 
  \scalebox{0.6}{\includegraphics{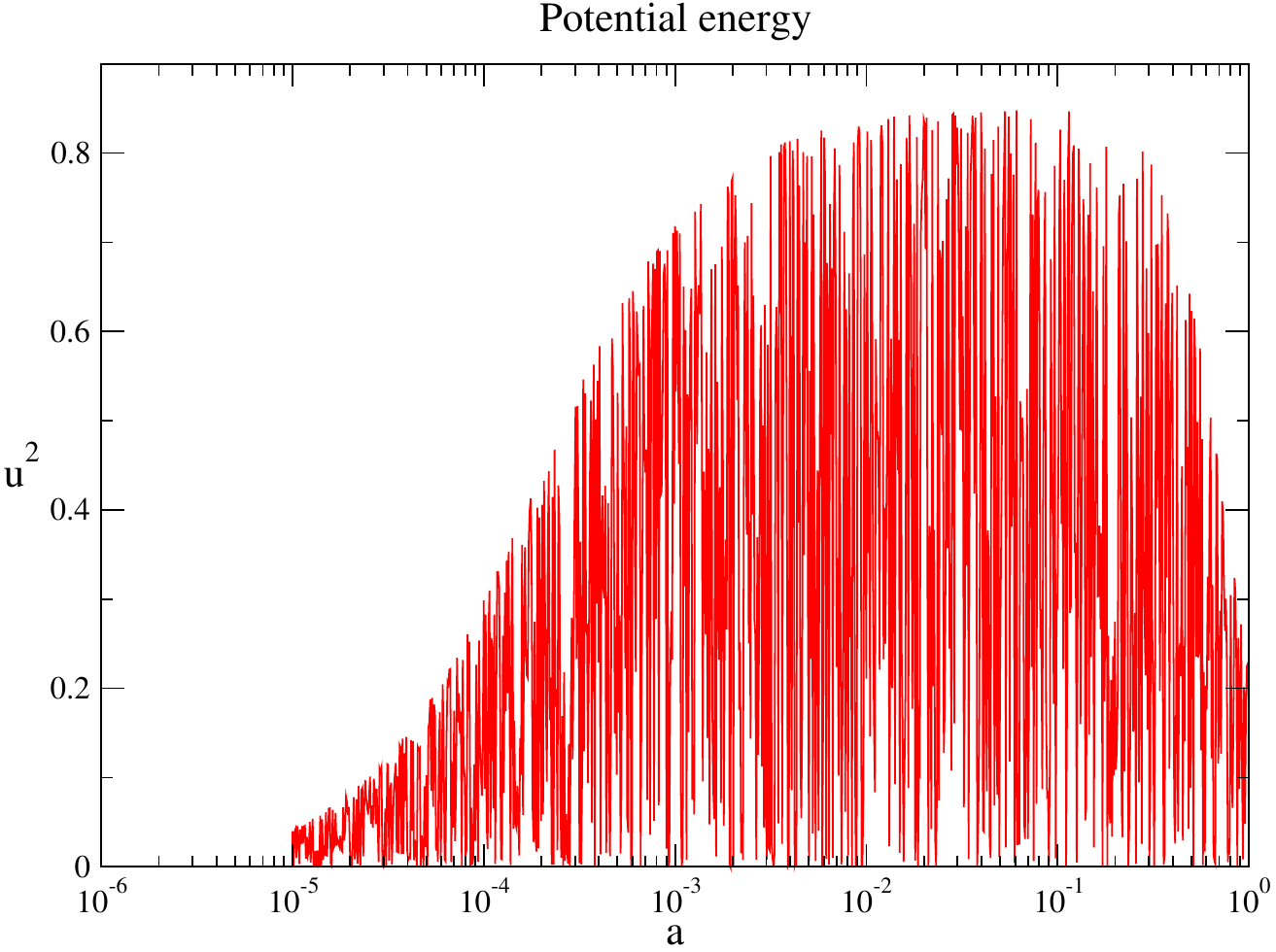}}
 \caption{Analytical evolution of the kinetic (top panel)  and the potential energy of the scalar field dark matter.}
\label{fig:analiticPot}
\end{figure}

Now we express the SF, $\Phi$, in terms of the new variables $S(t)$ and $\hat\rho(t)$, see \cite{suarez},
   \begin{equation}
    \Phi_0=2\sqrt{\hat{\rho}}\cos\left(S-m\,t\right).
   \label{tri}
   \end{equation}
This equation will allow us to give an analytical form to equation (\ref{rhophi0}). We first obtain
   \begin{equation}
\dot{\Phi}_0^2=\hat{\rho}\,\left[\frac{\dot{\hat{\rho}}}{\hat{\rho}}\cos\left(S- m\,t\right)
-2\left(\dot{S}-m\right)\,\sin\left(S-m\,t\right)\right]^{2}.
  \label{eq:tri}
   \end{equation}
We remind that we are working in coordinates such that $\vec{x}=a(t)\vec{R}$. 
This means that a structure in the Universe stand still
at point $\vec{R}$ and it is the space-time which is expanding. 
Using these coordinates for the
background, it follows $\dot{S}\sim0$ (since $S$ is related with the velocity). Since the background denisty 
is $(\ln\hat{\rho}_0)\dot{}=-3H$ with $H\sim 10^{-33}$eV$\ll m$, 
from (\ref{eq:tri}) we get
\begin{equation}
\dot{\Phi}_0^2=4m^2\hat{\rho}\sin^2\left(S-m\,t\right).
\label{eq:trif}
\end{equation}
Finally, substituting this last equation and equation (\ref{tri}) into (\ref{rhophi0}), we obtain,
   \begin{equation}
    \rho_{\Phi}=2m^2\hat{\rho}\left[\sin^2\left(S- m\,t\right)+\cos^2\left(S-m\,t\right)\right]=2m^2\hat{\rho.}
   \label{trigo}
   \end{equation}
Comparing this result with the first equation in (\ref{eq:dens}) we find that the identity $\Omega_{\Phi_0}=x^2+u^2=2m^2\hat{\rho}$ 
holds for the background. By comparing with (\ref{trigo}), we derive that
  \begin{eqnarray}
   x&=&\sqrt{2\,\hat\rho}\,m\sin\left(S-m\,t\right),\nonumber\\
   u&=&\sqrt{2\,\hat\rho}\,m\cos\left(S-m\,t\right).
   \label{eq:potencial}
   \end{eqnarray}
We plot the evolution of the potentials (\ref{eq:potencial}) in Fig.~\ref{fig:analiticPot}, in terms
of the e-folding number $N$ defined earlier and noting that 
$a\sim t^n$, implying $t\sim e^{N/n}$. 
%In terms of the two analytic results (\ref{eq:potencial}),  
Fig.~\ref{fig:analiticPot} shows the kinetic and the potential energies 
of the scalar field.
The match of the kinetic and potential energy of the background 
with the numerical results in Fig.~\ref{fig:Ox} is excellent.

\section{The Linear Regime of SF-dark matter perturbations}\label{sec:linear}

In this section we compute the growth of the SFDM overdensities $\delta\rho_{\Phi}$ in the linear regime.
%at the minimum of the potential where we have reached the $\Phi^2$ profile and we have the desired value for the mass $m$, 
In this regime, we suppose that the density contrast $\delta\equiv\delta\rho_{\Phi}/\rho_{\Phi_{0}}$ will be much smaller than unity. 
It is believed that the Universe was almost uniform after inflation, with a very small density contrast. 
As the Universe expanded, the small overdensities grew until they began to collapse, leading to the formation of structure in the Universe. 
Here we consider small deviations in the FLRW model, so that they can be treated by linear perturbation theory. 
After introducing the perturbed metric tensor in the FLRW background, we only consider scalar perturbations. 
We then give the equation of energy-momentum conservation and the Einstein field equations for the perturbed metric.

We separate the metric tensor as follows
\begin{equation}
g_{\mu\nu}=g_{\mu\nu}^0+\delta g_{\mu\nu},
\end{equation}
where $g_{\mu\nu}^0$ is the unperturbed metric for the background. Scalar perturbations can always be constructed by means of a scalar quantity, its derivatives, or any background quantity. We can then have a perturbed scalar metric to first order in terms of four scalars 
$\psi$ (lapse function), $\phi$ (gravitational potential), B (shift) and E (anisotropic potential), where
\begin{eqnarray}
\delta g_{00}&=&-a(\eta)^{2}2\psi, \nonumber\\
\delta g_{0i}&=&\delta g_{i0}=a(\eta)^{2}B,_i,\nonumber\\
\delta g_{ij}&=&-2a(\eta)^{2}(\phi\delta_{ij}-E,_{ij}),\label{eq:metric0}
\end{eqnarray}
with $\eta$ the conformal time.

From this, we get the most general perturbed line element
\begin{eqnarray}
ds^2&=&a(\eta)^2[-(1+2\psi)d\eta^2+2B,_id\eta dx^i\nonumber\\
&+&[(1-2\phi)\delta_{ij}+2E,_{ij}]dx^idx^j],\label{eq:metric1}
\end{eqnarray}
where $\delta_{ij}\mbox{ }$ is the background metric \cite{karim}. 
%and equations (\ref{eq:metric0})-(\ref{eq:metric1}) are given in conformal time $\eta$.

The perturbed energy-momentum tensor for the scalar field can be written as the background value $\mathbf{T}_0\equiv\mathbf{T}_0(t)$ plus a perturbation $\delta\mathbf{T}\equiv\delta\mathbf{T}(x^\mu)$ where $x^{\mu}=[t,x^i]$, i. e.
\begin{equation}
\mathbf{T}=\mathbf{T}_0+\delta\mathbf{T}.
\end{equation}

We consider the evolution equations within the Newtonian gauge, because it is a convenient gauge for the study of scalar perturbations. 
This gauge is defined when $B=0$ and $E=0$, and applies only to the scalar modes of the metric perturbations, implying that vector and tensorial modes 
are not taken into account.

We now derive the perturbed evolution equations for the different quantities mentioned above; the scalar perturbation $\delta\Phi$ 
and the scalar potential $\psi$. For the perturbed energy-momentum tensor, we have
\begin{subequations}
\begin{eqnarray}
\delta T^0_0&=&-\delta\rho_{\Phi}=-(\dot{\Phi}_0\dot{\delta\Phi}-\dot{\Phi}_0^2\psi+V,_{\Phi}\delta\Phi),\label{rhopert}\\
\delta T^0_i&=&-\frac{1}{a}(\dot{\Phi}_0\delta\Phi,_i),\\
\delta T^i_j&=&\delta P_{\Phi}=(\dot{\Phi}_0\dot{\delta\Phi}-\dot{\Phi}_0^2\psi-V,_{\Phi}\delta\Phi)\delta^i_j.
\end{eqnarray}\label{eq:Tperturbado}
\end{subequations}
In the above equations (\ref{eq:Tperturbado}) the dot denotes differentiating with respect to cosmological time $t$, 
which is related to the conformal time by $d/d\eta=a(d/dt)$.

In the Newtonian gauge, the metric tensor $g_{\mu\nu}$ becomes diagonal and 
therefore, 
the scalar potentials $\psi$ and $\phi$ are identical 
\begin{equation}
\psi-\phi=0.
\label{cond}
\end{equation}
We say that $\phi$ plays the role of the gravitational potential.
Usually this equation contains a term of anisotropic stress, which vanishes in the case of a scalar field. 
Altogether, to first order, the perturbed Einstein's equations $\delta G^i_j= \kappa^{2}\delta T^i_j$ for a scalar field 
in the Newtonian gauge are
\begin{eqnarray}
-8\pi G\delta\rho_{\Phi}&=&6H(\dot{\phi}+H\phi)-\frac{2}{a^2}\nabla^2\phi, \nonumber\\
8\pi G\dot{\Phi}_0\delta\Phi,_i&=&2(\dot{\phi}+H\phi),_i, \nonumber\\
8\pi G\delta P_{\Phi}&=&2[\ddot{\phi}+3H\dot{\phi}+(2\dot{H}+H^2)\phi],
\label{eq:sfmet}
\end{eqnarray}
which is in accordance with previous results obtained by \cite{karim,chung} and others. 
These equations describe the evolution of the scalar perturbations. 

Equations \eqref{eq:sfmet} can be rearranged to find an equation for 
$\phi$:
\begin{equation}
\ddot{\phi}+6H\dot{\phi}-\frac{1}{a^2}\nabla^2\phi+(2\dot H+4H^2)\phi+8\pi G\ V,_{\Phi} 
\delta\Phi=0.
\label{eq:gravPot}
\end{equation}
For the evolution of the perturbations in the scalar field we use the perturbed Klein-Gordon equation
\begin{equation}
\ddot{\delta\Phi}+3H\dot{\delta\Phi}-\frac{1}{a^2}\nabla^2\delta\Phi+V,_{\Phi\Phi}\delta 
\Phi+2V,_{\Phi}\phi-4\dot{\Phi}_0\dot{\phi}=0.
\label{eq:kgl}
\end{equation}

In order to solve equations (\ref{eq:sfmet}) and (\ref{eq:kgl}), we turn to Fourier's space.  
The usefulness of this expansion relies on the fact that each Fourier mode will propagate independently. 
To first order, the derivation of Fourier's components is straightforward. The perturbation $\delta\Phi$ is related to its Fourier component $\delta\Phi_k$ by
\begin{eqnarray}
\delta\Phi(t,x^i)&=&\int d^3k\delta\Phi(t,k^i)\,\textrm{exp}(ik_ix^i)\nonumber\\
&=&\int d^3k\delta\Phi_k\,\textrm{exp}(ik_ix^i),
\end{eqnarray}
where $k$ is the wavenumber. Here the wavenumber is defined as $k=2\pi /\lambda_{k}$, 
and $\lambda_{k}$ denotes the length scale of the perturbation (notice that $\lambda_{k}$ is different
to the self-interacting parameter $\lambda$).

The perturbed equations (\ref{eq:sfmet}) altogether with the scalar field read
\begin{subequations}
\begin{eqnarray}
8\pi G(3H\dot{\Phi}_0\delta\Phi_k)+\frac{2k^2}{a^2}\phi&=&-8\pi G(\dot{\Phi}_0\dot{\delta\Phi_k}\nonumber-
\phi\dot{\Phi}_0^2\\&&+V,_{\Phi}\delta\Phi_k),\label{eq:fsa}\\
2(H\phi+\dot{\phi})&=&8\pi G\dot{\Phi}_0\delta\Phi_k, \label{eq:fsb}\\
2[\ddot{\phi}+3H\dot{\phi}+(2\dot{H}+H^2)\phi]&=&
8\pi G(\dot{\Phi}_0\dot{\delta\Phi_k}-\phi\dot{\Phi}_0^2\nonumber\\
&&-V,_{\Phi}\delta\Phi_k)\label{eq:fsc}.
\end{eqnarray}
\label{eq:fsp}
\end{subequations}
For the corresponding Fourier transform of equation \eqref{eq:gravPot} we have,
\begin{equation}
\ddot{\phi_k}+6H\dot{\phi_k}+\left(\frac{k^2}{a^2}+2\dot H+4H^2\right)\phi_k+8\pi G\,V,_{\Phi} 
\delta\Phi_k=0.
\label{eq:gravPotk}
\end{equation}
and the Klein-Gordon equation (\ref{eq:kgl}) transforms into
\begin{equation}
\ddot{\delta\Phi}_k+3H\dot{\delta\Phi_k}+\left(\frac{k^2}{a^2}+V,_{\Phi\Phi}\right)\delta\Phi_k+
2\phi V,_{\Phi}-4\dot{\phi}\dot{\Phi}_0=0.
\label{eq:kgfou}
\end{equation}
These set of equations describe the evolution of the perturbations in the linear regime. Eq. (\ref{eq:fsa}) makes reference to the evolution 
of the energy density, Eq. (\ref{eq:fsb}) to the evolution of the gravitational potential and finally, Eqs. \eqref{eq:gravPotk} 
and \eqref{eq:kgfou} refer to the perturbations over the gravitational potential and the scalar field, respectively. 
Eqs. \eqref{eq:gravPotk} and \eqref{eq:kgfou} represent harmonic oscillators with a damping term that 
goes as $6H\dot{\phi_k}$ or $3H\dot{\delta\Phi_k}$, respectively, plus an extra force.
In Eq.~\eqref{eq:gravPotk} the term 2$\dot{H}$ is negative ($\dot{H}<0$), meaning that the gravitational 
fluctuations can grow up provided that the condition $\left(\frac{k^2}{a^2}+2\dot H+4H^2\right)<0$ is satisfied. 
In the scalar field perturbation (equation \eqref{eq:kgfou}), the situation is
a bit different because the perturbations 
can only grow if the condition $\left(\frac{k^2}{a^2}+V,_{\Phi\Phi}\right)<0$ is fullfiled. 
This can happen  if $V,_{\Phi\Phi}<0$ and $k^2/a^2$ is sufficiently small, which
implies that it is in a maximum of the potential, 
where the perturbation is unstable, and is rolling down to the minimum. 
Therefore the fluctuation can only grow up only during the phase transition of the scalar field, 
where the terms in the potential containing $\lambda$ cannot be neglected, because the critical temperature $T_c$ has not been reached. 
After that, the scalar fluctuations will continue to grow up and therefore, they could form gravitational structures.

Now, taking the time derivative of (\ref{rhopert}), we get
\begin{equation}
 \dot{\delta\rho_{\Phi}}=(\ddot{\Phi}_0+V,_{\Phi})\dot{\delta\Phi}
+(\ddot{\delta\Phi}+V,_{\Phi\Phi}\delta\Phi-\dot{\Phi}_0\dot{\phi})\dot{\Phi}_0
-2\phi\dot{\Phi}_0\ddot{\Phi}_0.
\end{equation}
Performing a Fourier transformation to the above equation and using 
%the unperturbed and perturbed Klein-Gordon 
equations (\ref{eq:back}), (\ref{eq:fsp}) and (\ref{eq:kgfou}), we arrive at
\begin{equation}
\dot{\delta\rho_{\Phi}}=-6H\dot{\Phi}_0\dot{\delta\Phi_k}+6\phi_k\dot{\Phi}_0^2H-\frac{2k^2}{a^2\kappa^2}(H\phi_k+\dot{\phi}_k)+3\dot{\phi}_k
\dot{\Phi}_0^2.
\end{equation}
On the other hand we have
\begin{equation}
\delta P_{\Phi}+\delta\rho_{\Phi}=2\dot{\Phi}_0\dot{\delta\Phi_k}-2\dot{\Phi}_0^2\phi_k,
\end{equation}
Thus,
\begin{equation}
\dot{\delta\rho_{\Phi}}=-3H(\delta P_{\Phi}+\delta\rho_{\Phi})-\frac{2k^2}{a^2\kappa^2}(H\phi_{k}+\dot{\phi_{k}})+3\dot{\phi_{k}}\dot{\Phi}_0^2.
\end{equation}
Since we know that 
\begin{equation}
\dot{\rho_{\Phi_{0}}}=-3H(\rho_{\Phi_{0}}+ P_{\Phi_{0}}).
\end{equation}
we obtain
\begin{equation}
\dot\delta+3H\left(\frac{\delta P_{\Phi}}{\delta\rho_{\Phi}}-\omega_{\Phi_{0}}\right)\delta=3\dot{\phi_{k}}\left(1+\omega_{\Phi_{0}}\right)-G_{\phi},
\label{pro1}
\end{equation}
where we have defined the function $G_{\phi}$ as
\begin{equation}
G_{\phi}=\frac{2k^2}{a^2\kappa^2}\frac{\dot{\phi_{k}}+H\phi_{k}}{\rho_{\Phi_{0}}}.
\end{equation}
Taking the time average of equation (\ref{pro1}) we obtain
\begin{equation}
\dot\delta+3H\left(\left<\frac{\delta P_{\Phi}}{\delta\rho_{\Phi}}\right>-\left<\omega_{\Phi_{0}}\right>\right)\delta=3\dot{\phi_{k}} 
\left<F_{\Phi}\right> -\left<G_{\phi}\right>.\label{eq:deltlini}
\end{equation}  
where $F_{\phi}$ is defined as 
\begin{equation}
F_{\Phi}=1+\omega_{\Phi_{0}}.
\end{equation}
In the radiation and matter dominated eras, the term
$\left<\delta P_{\Phi}/\delta\rho_{\Phi}\right>$ in
Eq.~(\ref{eq:deltlini}) is $\approx 0$, 
see for example \cite{matosurena00b}. 
This is because $\delta P_{\Phi}$ oscillates very rapidly
around zero, whereas $\delta \rho_{\Phi}$ stays almost constant
during one oscillation of $\delta P_{\Phi}$. This behaviour 
will be confirmed numerically in \S \ref{sec:numreslr}. 
Note however, that 
the time average $\left<\frac{\delta P_{\Phi}}{\delta \rho_{\Phi}}\right>$ 
might not necessarily be identical to zero.
Also we can see from Fig.~\ref{fig:w0} that 
$\left<\omega_{\Phi_{0}}\right>\approx0$. Moreover, since we are using 
post-Newtonian approximation, 
$G_{\phi}$ can be neglected, as the numerical results will confirm later on. 
Consequently, not only 
the scalar field $\Phi$ behaves very similar 
as the $\Lambda$CDM model in the background Universe,
but equation (\ref{eq:deltlini}) tell us that their perturbations do too; 
the growing behavior for the $k$ modes are recovered and preserved so far.

As a final remark, we want to point out that in the case that the perturbation $\delta\Phi (\vec{x},t)$ fullfills equation (\ref{tri}), 
but now with $S(\vec{x},t)$ and $\hat\rho(\vec{x},t)$ also depending on position, then the perturbed Klein-Gordon equation (taking $\dot\phi =0$ for simplicity) can be rewriten as (see \cite{suarez}):
 \begin{eqnarray}
  -\dot{S} &+&\frac{{\dot S}^2}{2m}+\frac{1}{2m}\frac{\Box\sqrt{\hat\rho}}     
  {\sqrt{\hat\rho}}-\frac{(\nabla S)^2}{2ma^2}
  -m\phi +\frac{9}{2}\frac{\lambda}{m}\hat{\rho}=0,\nonumber\\ 
  {\dot{\hat\rho}} &+& 3H\hat\rho+\frac{1}{m}(\hat\rho\Box S +\frac{1}{a^2}\nabla{S}\nabla\hat{\rho}-\dot S{\dot{\hat\rho}})=0,
   \end{eqnarray}
 where the D'Alambertian operator $\Box$ is now given by $\Box=-\partial^2_t-3H\partial_t+\nabla^2/a^2$ and $\lambda\neq 0$ for the moment. 
From this set of equations we get the following constraint equation in the Fourier space
$$\mathrm{w}^2=(v_q^2+\mathfrak{w}\rho_0)\frac{k^2}{a^2}-4\pi G\rho_0,$$
where 
$\mathfrak{w}=-9\lambda/2m^2$ and
$v^2_q=k^2/4m^2$
is called the quantum velocity (in this case it is not the velocity of sound) and it is associated with the nature of quantum fluctuations. 
$\rho_0$ is the background density,
so that $\rho_0\sim 1/a^3$.
We may write this equation as
\begin{equation}
\mathrm{w}^2=4\pi^2(v_q^2+\mathfrak{w}\rho_0)\left[\frac{1}{a^2\lambda_{k}^2}-\frac{G\rho_0}{\pi(v_q^2+\mathfrak{w}\rho_0)}\right],
\end{equation}
which defines the wavelength $\lambda_{kJ}\equiv\sqrt{\pi\left(v_q^2+\mathfrak{w}\rho_0\right)/G\rho_0 a^2}$. This equation  relates the wavelength 
of the perturbation to the size of the gravitational structure to be formed in a Universe in expansion for the model $\lambda\Phi^4$. 
In the limit $\lambda\sim 0 $ (when we have reached the minimum and we have a $\Phi^2$ profile), then 
$\lambda_{kJ}=\sqrt{\pi v_q^2/G\rho_0a^2}$, which is
the Jeans length in a Universe in expansion, 
with quantum fluctuations as the origin of structure formation.
This expression is analogous to the effective Jeans length for a scalar field \citep{chung2,hwang,mota_bruck,khlopov} given by
\begin{equation}
\lambda_{kJ}\approx 2\pi/\sqrt{V,_{\Phi\Phi}}\approx2\pi/m.
\end{equation}
For instance, for an ultralight mass $\sim10^{-23}$ eV, the Jeans length in the recombination era is $\lambda_{kJ}\sim4$ kpc. Therefore, 
on scales much smaller than this value, the scalar perturbations will not grow and they will not form gravitational structures.
It is worth to note that once the value of the scalar mass is fixed, 
a natural cut in the mass power spectrum arises in a natural way. Therefore, 
the SFDM model avoids the problem of excessive substructure in the 
$\Lambda$CDM \citep{hu,matosurena00b}. 

\subsection{Numerical results for SFDM perturbations in the linear regime for $T\sim T_{c}$}
In this section we perform a numerical study of the system of equations 
(\ref{eq:back}), (\ref{eq:gravPotk}) and (\ref{eq:kgfou}) at an early stage of the Universe where 
the finite temperature effects are very relevant for the evolution of SFDM/BEC perturbations as well as for the
density profile of SFDM/BEC halos \citep{sle11, harkomada, harkomocanu}.
In this temperature regime, $T \sim T_{c}$, the scalar potential takes the form  $V(\tilde{\Phi})=-\frac{1}{2}\mathfrak{m}^{2}\tilde\Phi^{2} + \frac{\lambda}{4}\tilde\Phi^{4}+
\frac{\lambda}{8}T^2\tilde\Phi^2-\frac{\pi}{90}T^4+\frac{\mathfrak{m}^4}{4\lambda}$.
\begin{figure}[ht]
 \centering
 \scalebox{0.6}{\includegraphics{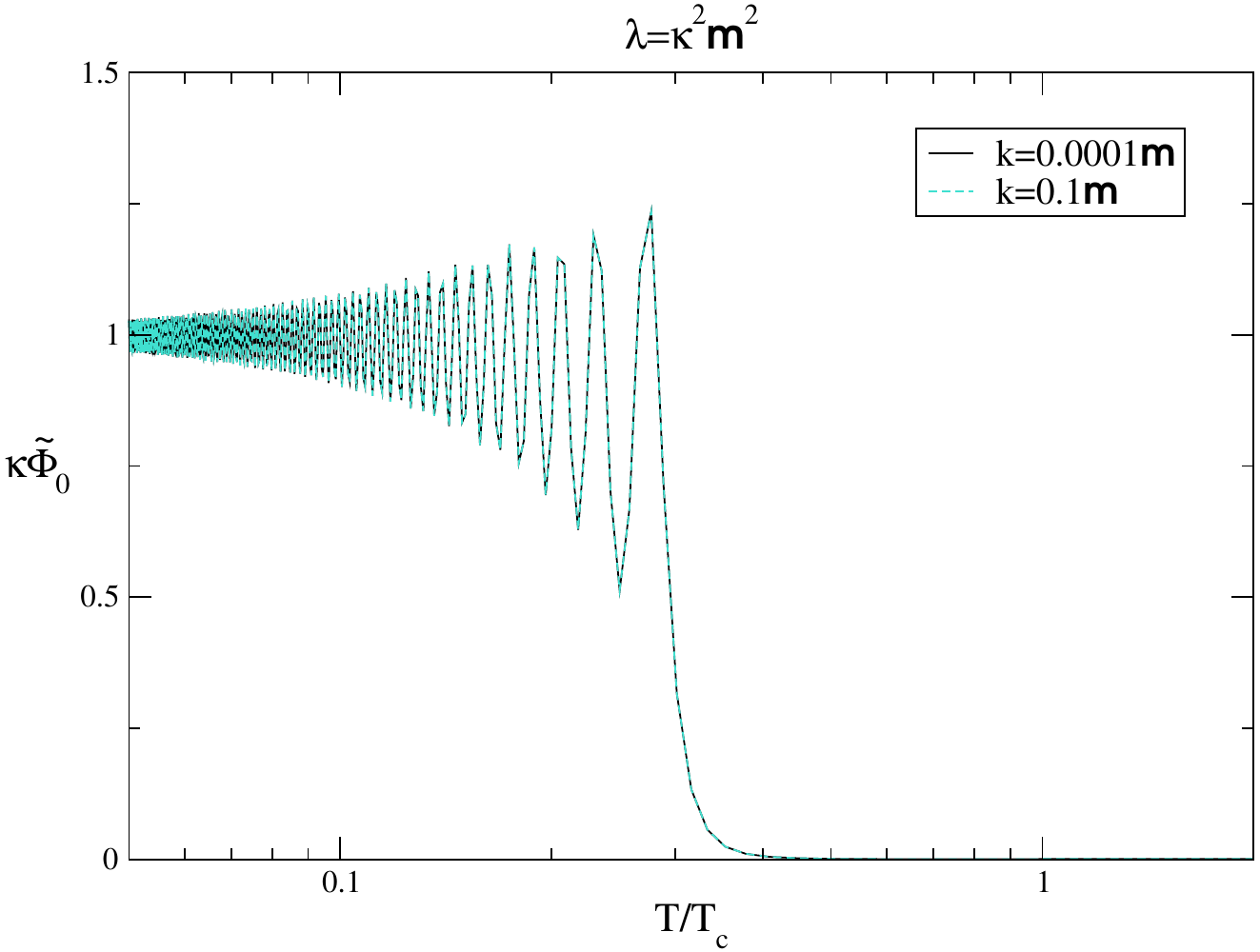}}
 \scalebox{0.6}{\includegraphics{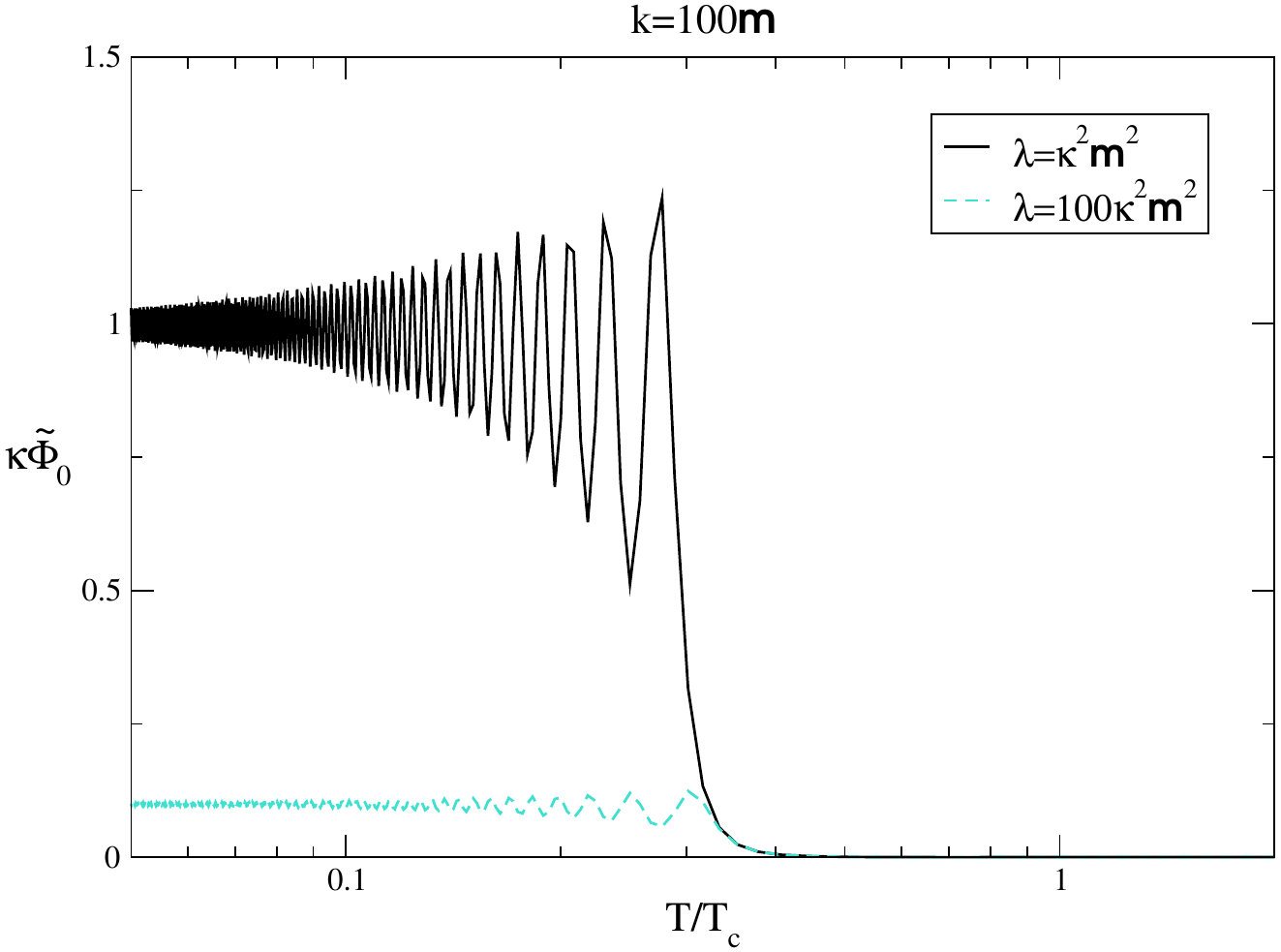}}
 \caption{Evolution of the scalar field $\tilde{\Phi}_0(t)$ in the temperature regime $T\sim T_{c}$ with the scalar potential (\ref{mexicanhat}).}
%in the $-\frac{1}{2}\mathfrak{m}^2\Phi^{2}+\frac{\lambda}{4}\Phi^4+\frac{\lambda}{8}\Phi^2 T^2+\frac{\pi}{90} T^4+\frac{\mathfrak{m}^4}{4\lambda}$-dark matter model.}
 \label{fig:phi_T}
\end{figure}
The numerical results in Fig.~\ref{fig:phi_T} are obtained for the potential (\ref{mexicanhat}). We have also used equation (\ref{eq:back}) 
(at very early epochs, only the radiation and scalar fields need to
be considered)
which is an expression valid for any SF potential. 
At temperatures close to $T_c$ we are not yet at the minimum of the scalar potential 
(the mass term is negative, and $\tilde\Phi_0(t)$ has not found its minima), and we are at a point near the phase transition of the scalar field. 
Fig.~\ref{fig:phi_T} shows two interesting cases. First, we keep the self-interacting parameter $\lambda$ fixed and we take different values for the wavenumber $k$ (top panel).
In the bottom panel, we show the evolution of the scalar field $\tilde{\Phi}_0(t)$ for different values of $\lambda$. 
As we can see, the scalar field $\tilde{\Phi}_0$ has the same behavior for several values of $k$ (top panel). However, the oscillations of $\tilde{\Phi}_0$ 
have different amplitudes for several values of $\lambda$.
As mentioned before, at points near $T/T_c=1$ there is a sudden change on the value of $\tilde\Phi_0$, possibly associated with the symmetry breaking phase transition. 
After this jump when $T<T_c$ and the SF is searching for its minima, we can see that, at some point ($T/T_c\sim 0.1$ onwards), the SF stabilizes 
and oscillates around the same value. We believe that it is at this stage that the SF has found one of its stable minima and, 
from here on, all of our previous results may be applied.

\begin{figure}[ht]
 \centering
 \scalebox{0.6}{\includegraphics{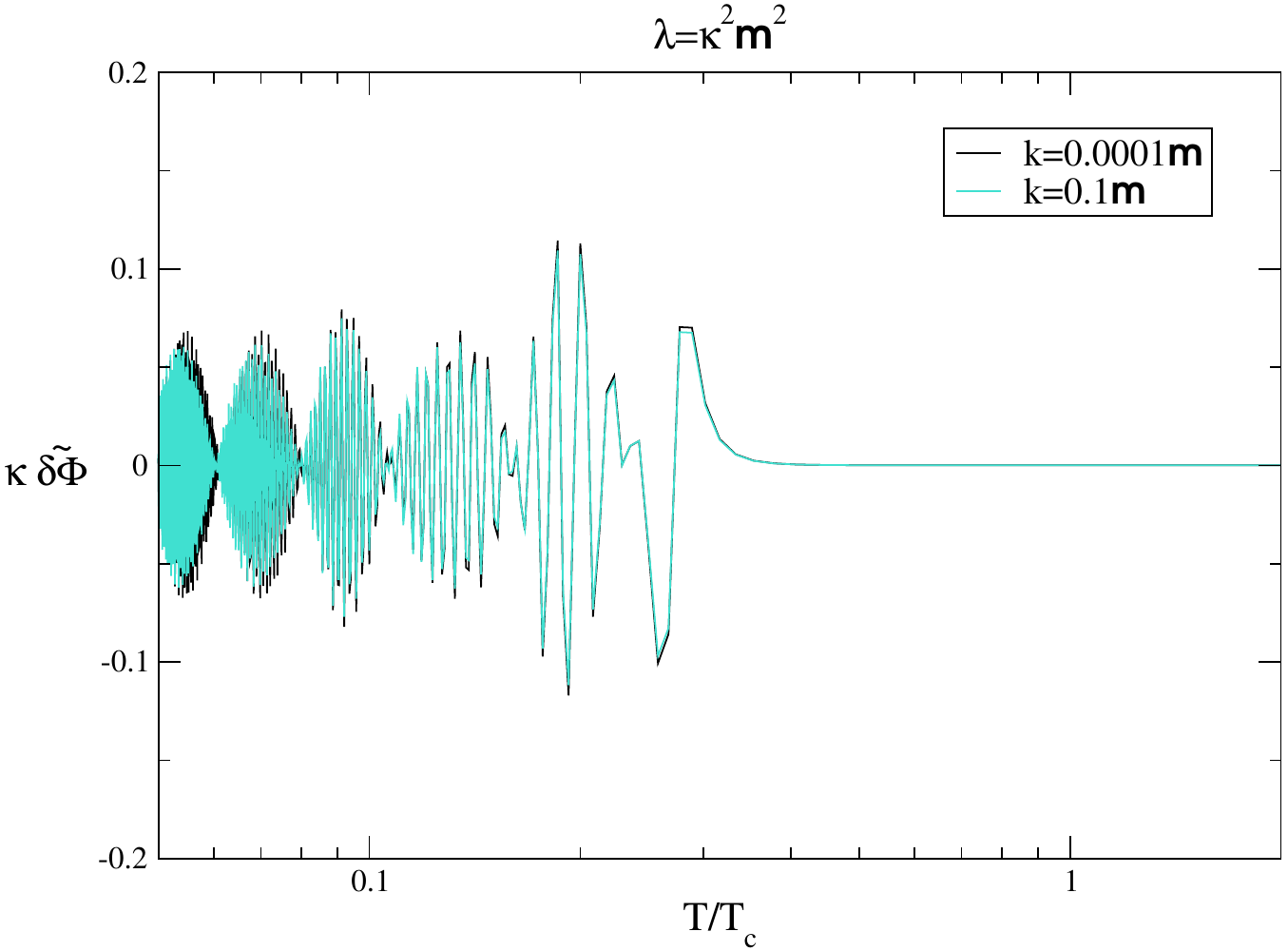}}
\scalebox{0.6}{\includegraphics{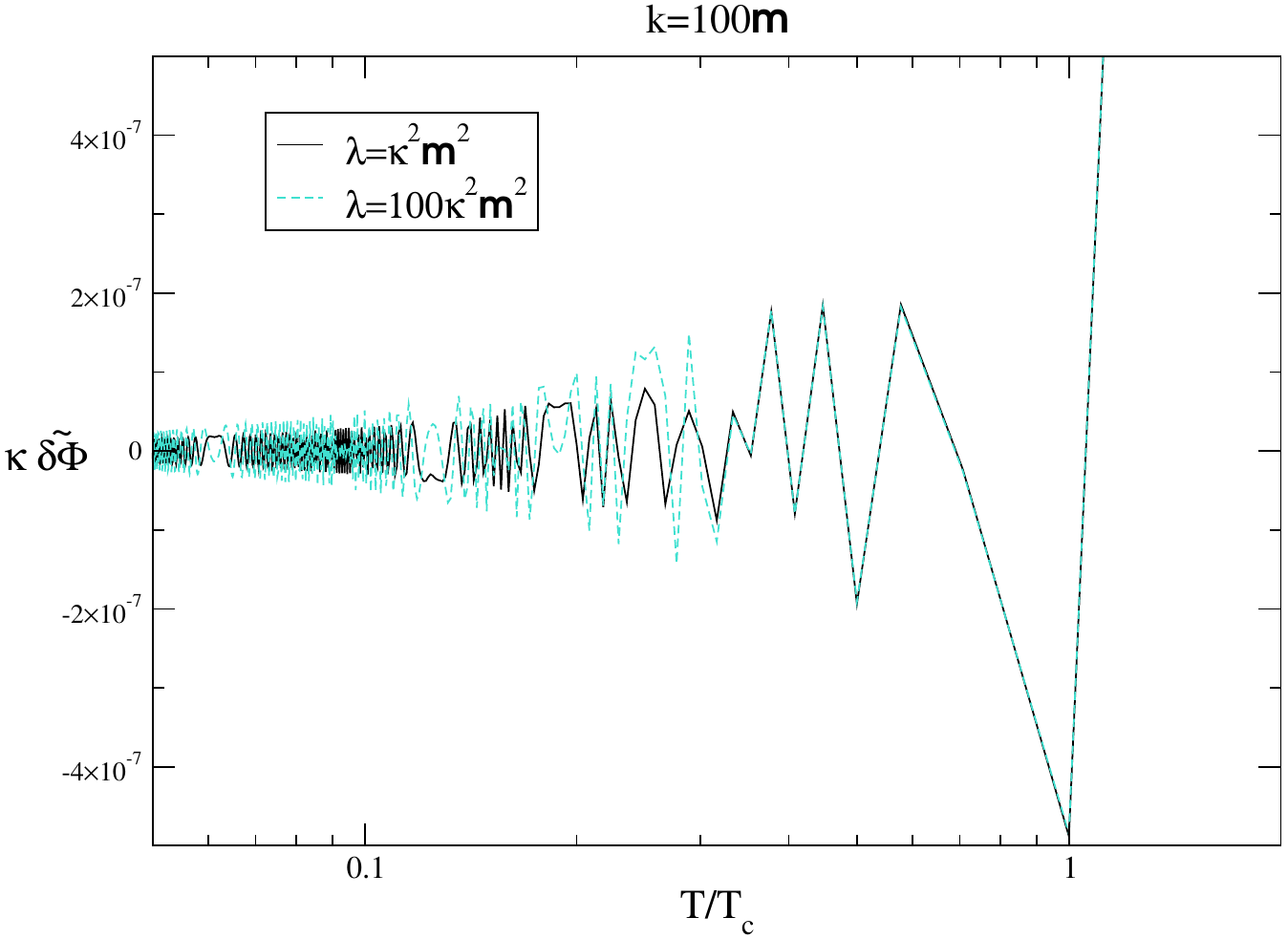}}
 \caption{Evolution of the scalar perturbation $\delta\tilde{\Phi}(\vec{x},t)$ in the temperature regime $T\sim T_{c}$ with the scalar potential (\ref{mexicanhat}).}
%$-\frac{1}{2}\mathfrak{m}^2\Phi^{2}+\frac{\lambda}{4}\Phi^4+\frac{\lambda}{8}\Phi^2 T^2+T_0 T^4+\frac{\mathfrak{m}^4}{4\lambda}$-dark matter model, corresponding to a 200Mpc perturbation.}
 \label{fig:delphi_T}
\end{figure}

In Fig.~\ref{fig:delphi_T} we show the evolution of the SF perturbation, $\delta\tilde{\Phi}(\vec{x},t)$, obtained through the numerical study of equation (\ref{eq:kgfou})
with the scalar potential (\ref{mexicanhat}) (recall that the mass term with $\mathfrak{m}$ has not reached its postive value at the minimum). 
Again, we explore the same two cases as before, 
we keep the self-interacting parameter $\lambda$ fixed and take different values for the wavenumber $k$ (top panel). 
We also show the evolution for different values of $\lambda$
(bottom panel). Observe that the amplitude of the oscillations of $\delta\tilde{\Phi}(\vec{x},t)$ 
drops slightly for bigger values of $k$ (top panel). 
However, the amplitude of the $\delta\tilde{\Phi}$ perturbations
increases with $\lambda$.
The results show that at points where $T<T_c$, the fluctuations grow 
(top panel). This growth starts just at $T=T_c$, 
at the phase transition point, with $\delta\tilde{\Phi}$ always oscillating around zero, taking very small values, as mentioned before. 
This result might give some insight on the relation between the growth of perturbations of the SF due to the phase transition with the growth of such perturbations
related to the formation of structures in the Universe.

\subsection{Numerical results for SFDM perturbations in the linear regime for $T\ll T_{c}$}\label{sec:numreslr}
\begin{figure}[ht]
 \centering
 \scalebox{0.6}{\includegraphics{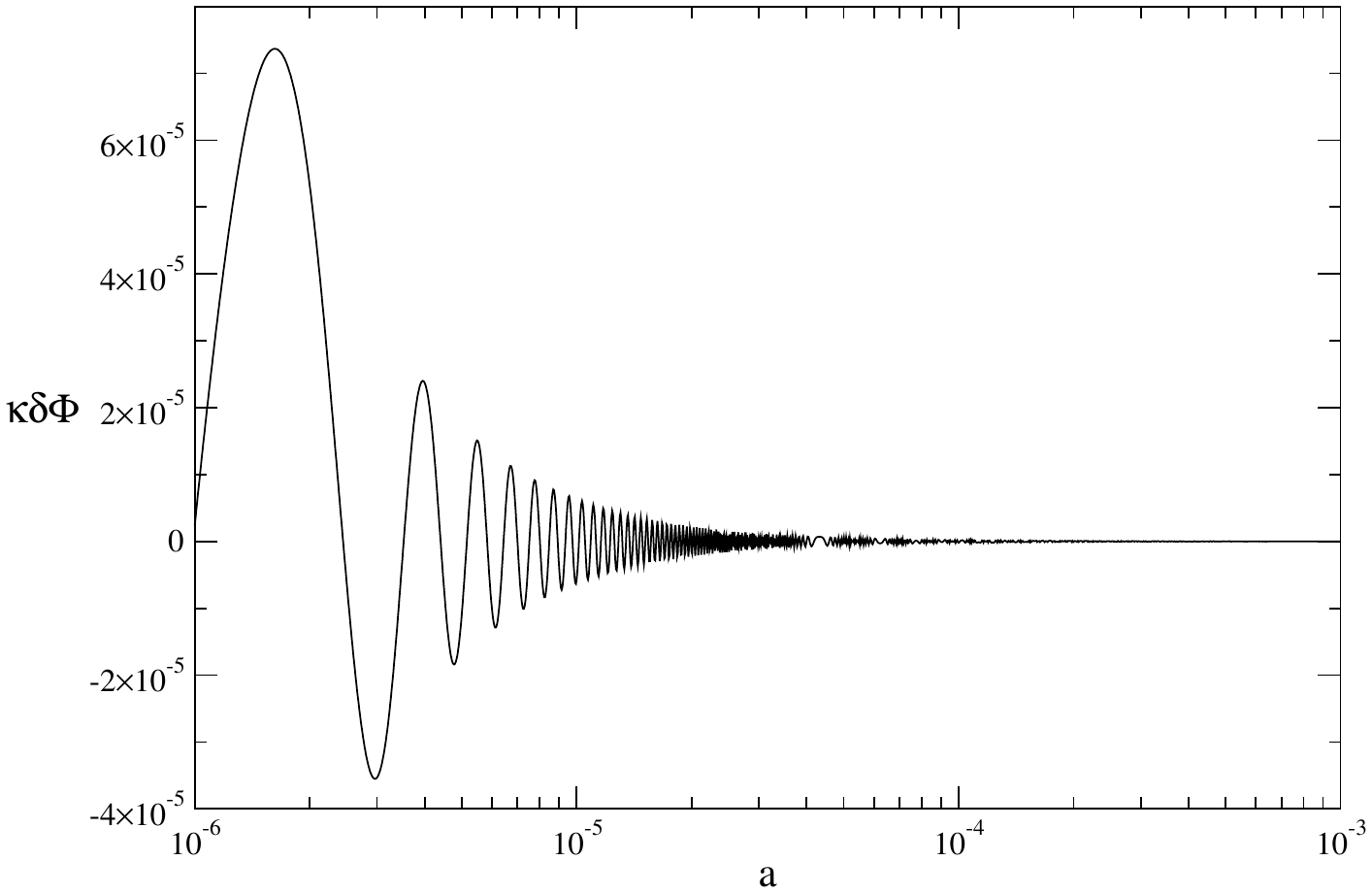}}
 \caption{Evolution of the perturbed scalar field $\delta\Phi$ as a function of the scale factor $a$.}
 \label{fig:delPhi}
\end{figure}

\begin{figure}[ht]
 \centering
 \scalebox{0.6}{\includegraphics{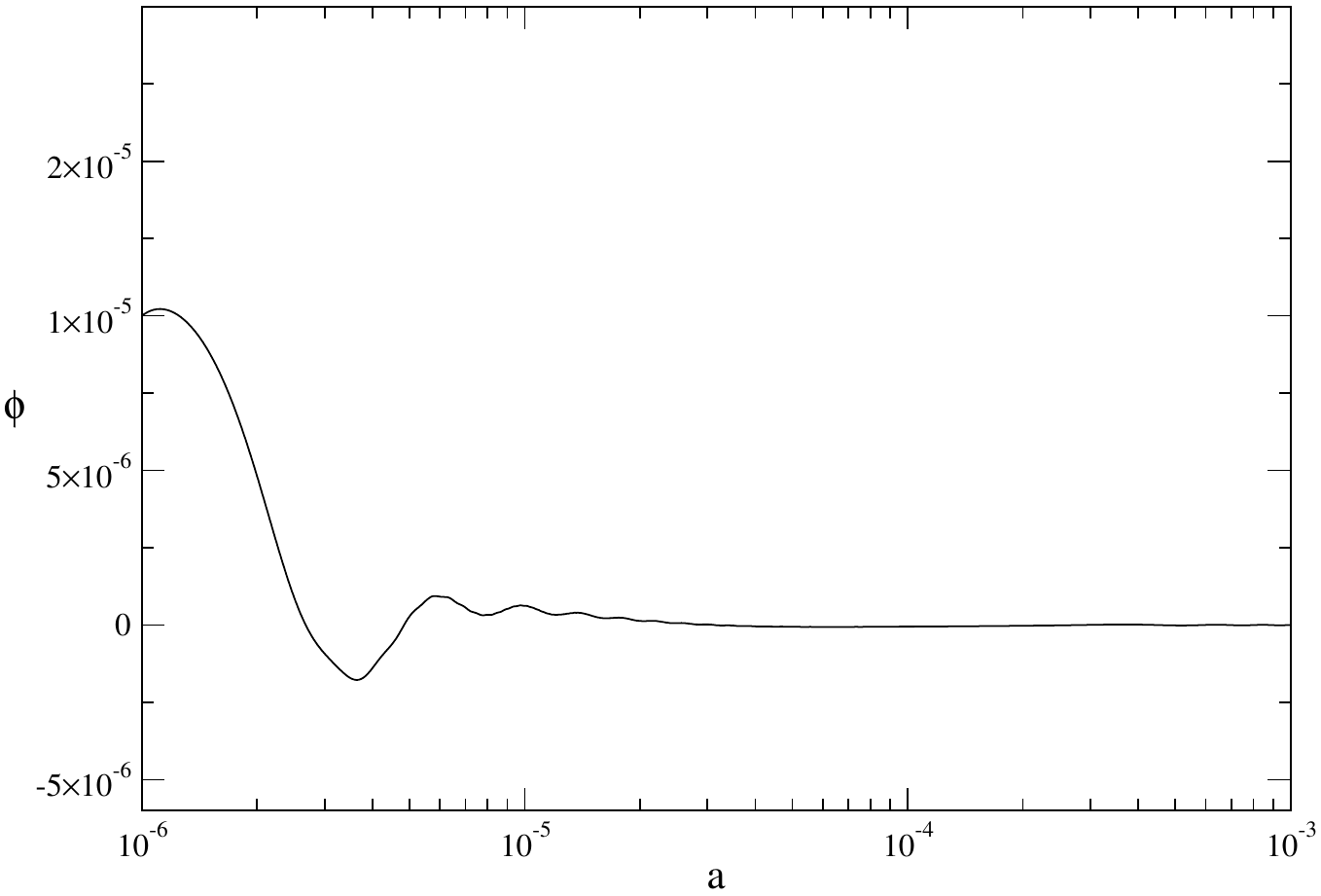}}
 \caption{Evolution of the gravitational potential $\phi$ as a function of the scale factor $a$.}
 \label{fig:gravpot}
\end{figure}

In this section, we will study the cosmological evolution of the growth of SF overdensities, $\delta\rho_{\Phi}$, in the linear regime when (1)
the temperature is $T\ll T_{c}$,
(2) the SF has reached the minimum of the potential with a $\Phi^{2}$-profile 
and (3) the mass term is positive and is described with $m$ 
\footnote{See \cite{boyle,johnson} for interesting results of the growth of scalar perturbations in this regime.}.
In order to obtain a numerical solution for the density contrast $\delta=\delta\rho_\Phi/\rho_{\Phi_0}$, 
the following dimensionless variables are defined,
%we make use of equations (\ref{rhophi0}) and (\ref{rhopert}). 
\begin{eqnarray}
l_{1}&\equiv&\phi_{k},\qquad\qquad
l_{2}\equiv\dot{\phi_{k}}/H,\qquad\qquad y_{1}=\delta,\nonumber\\
z_{1}&\equiv&\frac{\kappa}{\sqrt{6}}\delta\Phi_k,\quad\,
z_{2}\equiv\frac{\kappa}{\sqrt{6}}\frac{\delta\dot{\Phi}_k}{H}.
\label{eq:dvlr}
\end{eqnarray}
Using these variables, equations (\ref{eq:gravPotk}) and (\ref{eq:kgfou}) can be transformed into 
an autonomous dynamical system with respect to the e-folding number
\begin{subequations}
\begin{eqnarray}
l_{1}'&=&l_{2}, \\
l_{2}'&=&3l_{2}\left(\frac{\Pi}{2}-2\right) + l_{1}\left(3\Pi - 4\right)
-6 z_{1}\,u\,s -\frac{k^{2}s^{2}l_{1}}{m^{2}a^{2}},  \\
z_{1}'&=&z_{2}, \\
z_{2}'&=&3 z_{2} \left(\frac{\Pi}{2}-1\right) -z_{1}\,s^{2}\left(\frac{k^2}{a^{2}\,m^{2}} + 1\right)-2u\,s\,l_{1}+ 4l_{2}x,\\
y_{1}'&=& -3\left[\left(\frac{xz_{2}-x^{2}l_{1}-usz_{1}}{xz_{2}-x^{2}l_{1}+usz_{1}}\right)-\omega_{\Phi_{0}}\right]y_{1}+3l_{2}F_{\Phi}-\frac{G_{\Phi}}{H}\label{eq:deltanum},
\end{eqnarray}
\label{eq:dslin}
\end{subequations}
where $G_{\Phi}/H=2k^{2}s^{2}\left(l_{1}+l_{2}\right)/ 3a^{2}m^{2} \Omega_{\Phi_{0}}$ and
the functions $s$, $x$, $u$, and $\Pi$ were determined in section \ref{sec:background}.
Instead of using equation (\ref{eq:deltanum}),
which is a differential equation
for the linear density contrast $y_{1}$,
it is possible to obtain an algebraic expression for $y_{1}$ using
equations (\ref{rhophi0}) and (\ref{rhopert}),
which in terms of the dimensionless variables defined before
can be written as
\begin{equation}
y_{1}=\frac{2\left[x\left(z_2-xl_1\right)+ u s z_{1}\right]}{\Omega_{\Phi_{0}}}.
\label{eq:deltafor}
\end{equation}
A numerical solution for the system of equations (\ref{eq:dslin}) was obtained. 
We pose the initial conditions at $a_{i}=10^{-6}$. We start with a perturbation with wavelength $\lambda_{k}=2$ Mpc 
and density contrast $\delta=1\times10^{-7}$. Fig. \ref{fig:delPhi} shows the cosmological evolution of the perturbed scalar field as a function of the scale factor $a$. 
As we said before at very early epochs of the Universe the SF was in thermal equilibrium with its surroundings and the temperature of the Universe was very high, 
dominated mainly by radiation, thus making the amplitude of the fluctuations of the SF very large due to its interactions. 
\begin{figure}[ht]
 \centering
 \scalebox{0.6}{\includegraphics{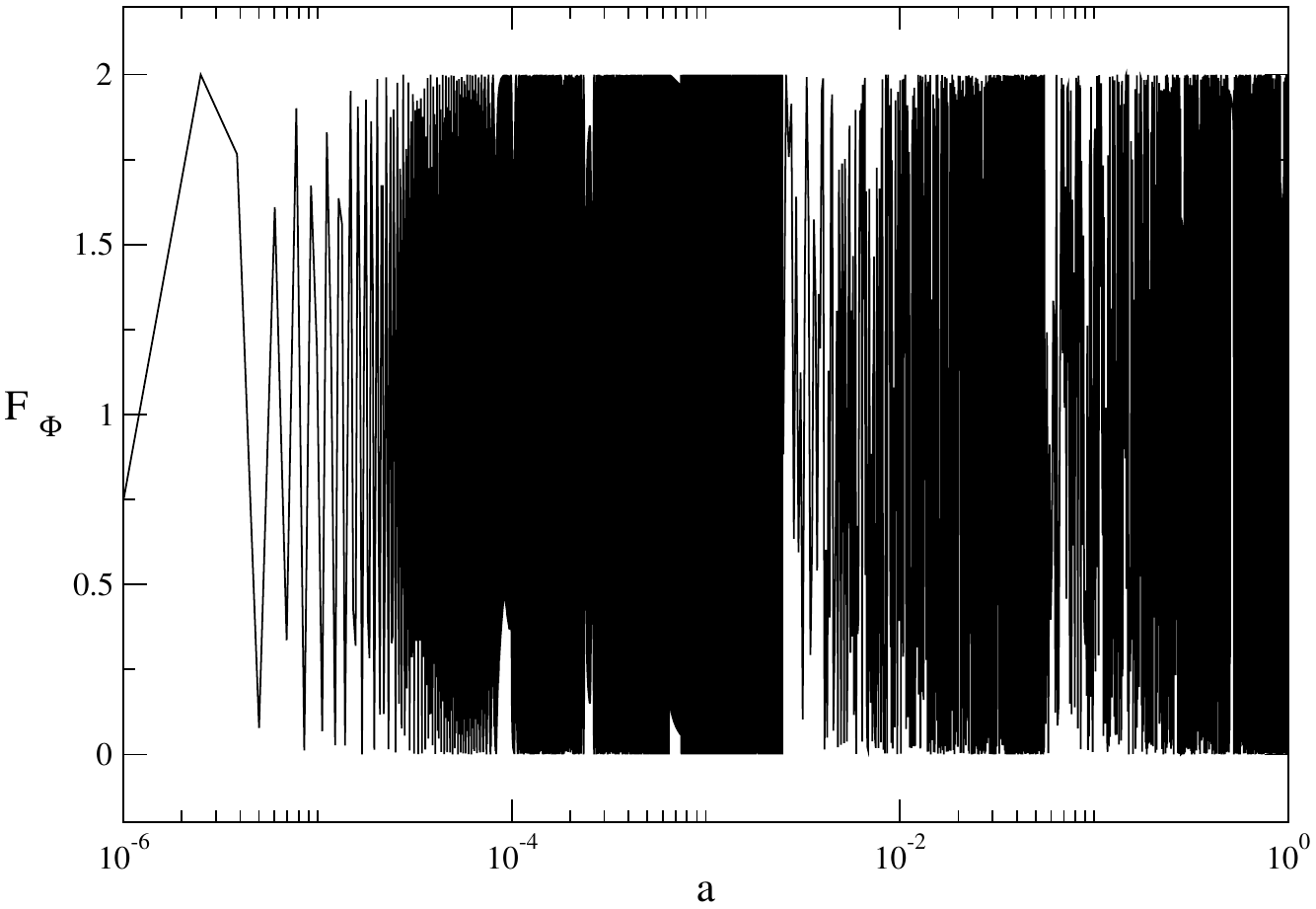}}
 \caption{Evolution of $\left< F_{\Phi} \right>$ term involved on the right-hand side of equation (\ref{eq:deltlini}).}
 \label{fig:FPhi0}
\end{figure}
As the temperature decreases the SF decouples from the rest of the matter, so that the surrounding interactions are negligible
after the decoupling. 
After the breaking of symmetry, when $T<T_c$, the SF begins to oscillate uniformly on the space surrounding its true minima, where the perturbations seem to stabilize, around $a\sim 10^{-4}$.  
The evolution of the gravitational potential for this perturbation is shown in Fig.~\ref{fig:gravpot}. Note that the gravitational potential remains constant from $a\sim10^{-5}$ all along up to 
the matter-dominated regime.

\begin{figure}[ht]
 \centering
 \scalebox{0.6}{\includegraphics{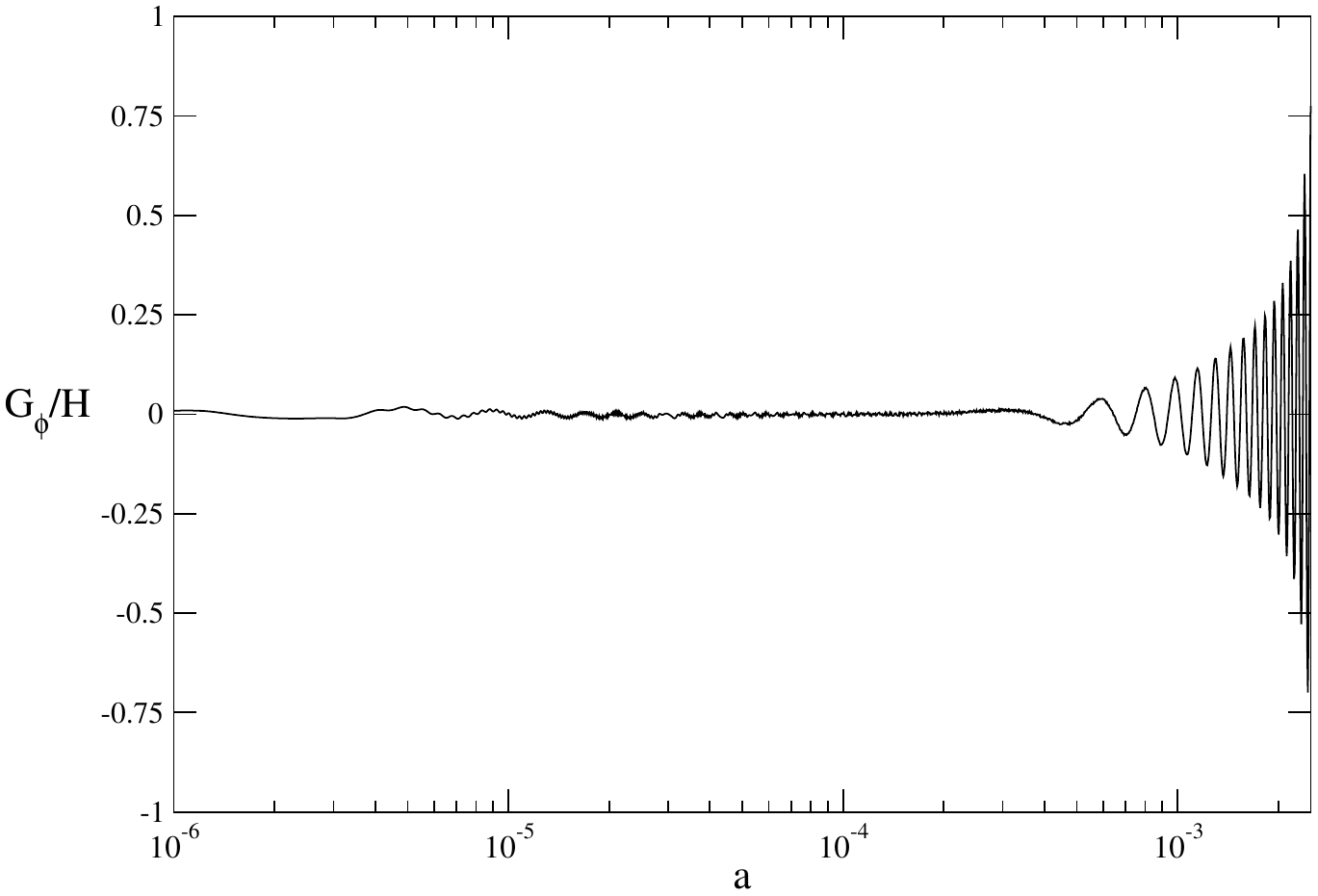}}
 \caption{Evolution of $\left< G_{\phi} \right>$ term involved on the right-hand side of equation (\ref{eq:deltlini}).}
 \label{fig:GPhi}

\end{figure}

\begin{figure}[ht]
 \centering
 \scalebox{0.6}{\includegraphics{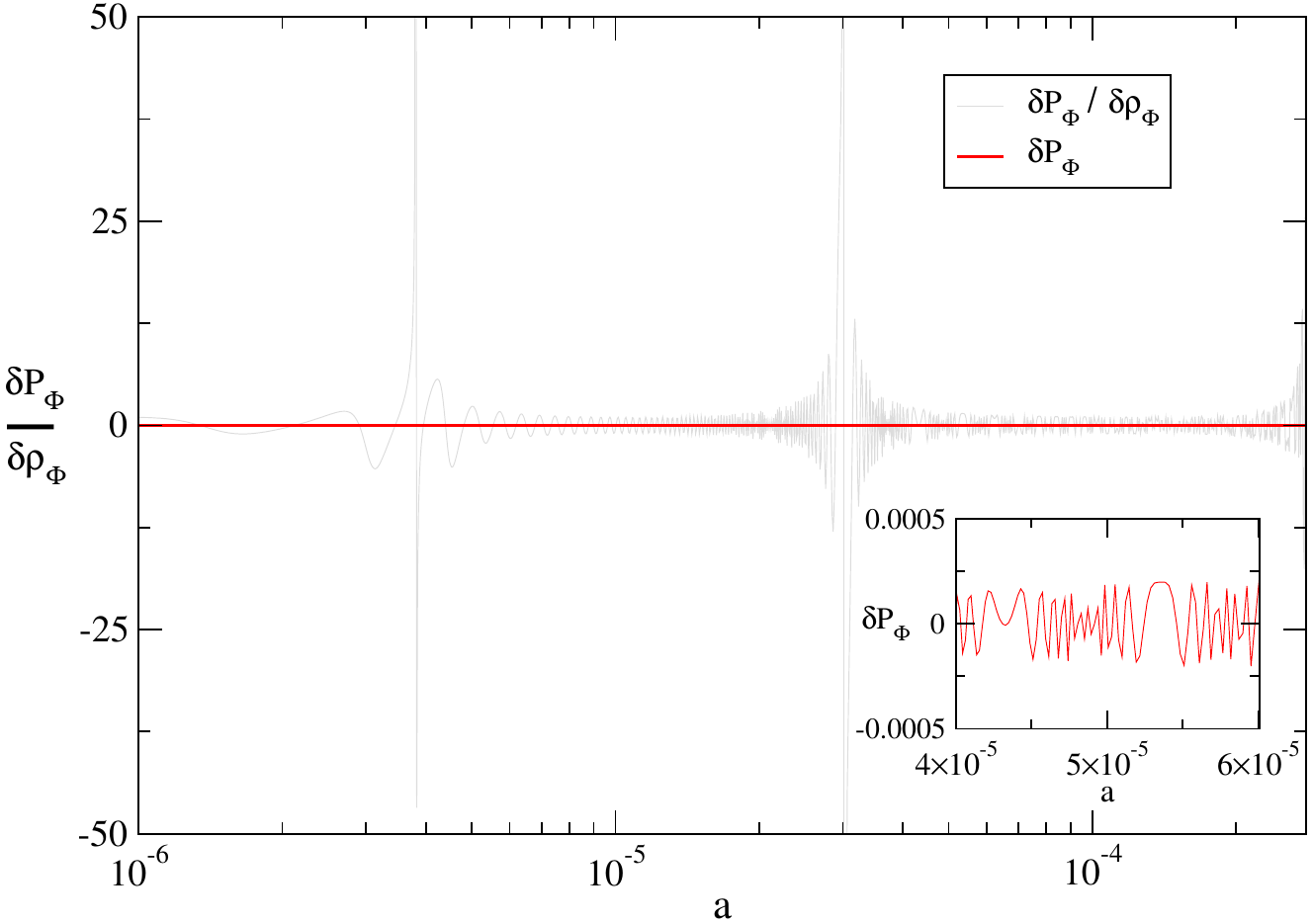}}
 \caption{Evolution of $\delta{P_{\Phi}} / \delta{\rho_{\Phi}} $ term involved on the left-hand side of equation (\ref{eq:deltlini}).
The inset shows the evolution of $\delta{P_{\Phi}}$ in a short interval of $a$.}
 \label{fig:dp}
\end{figure}

\begin{figure}[ht]
 \centering
 \scalebox{0.6}{\includegraphics{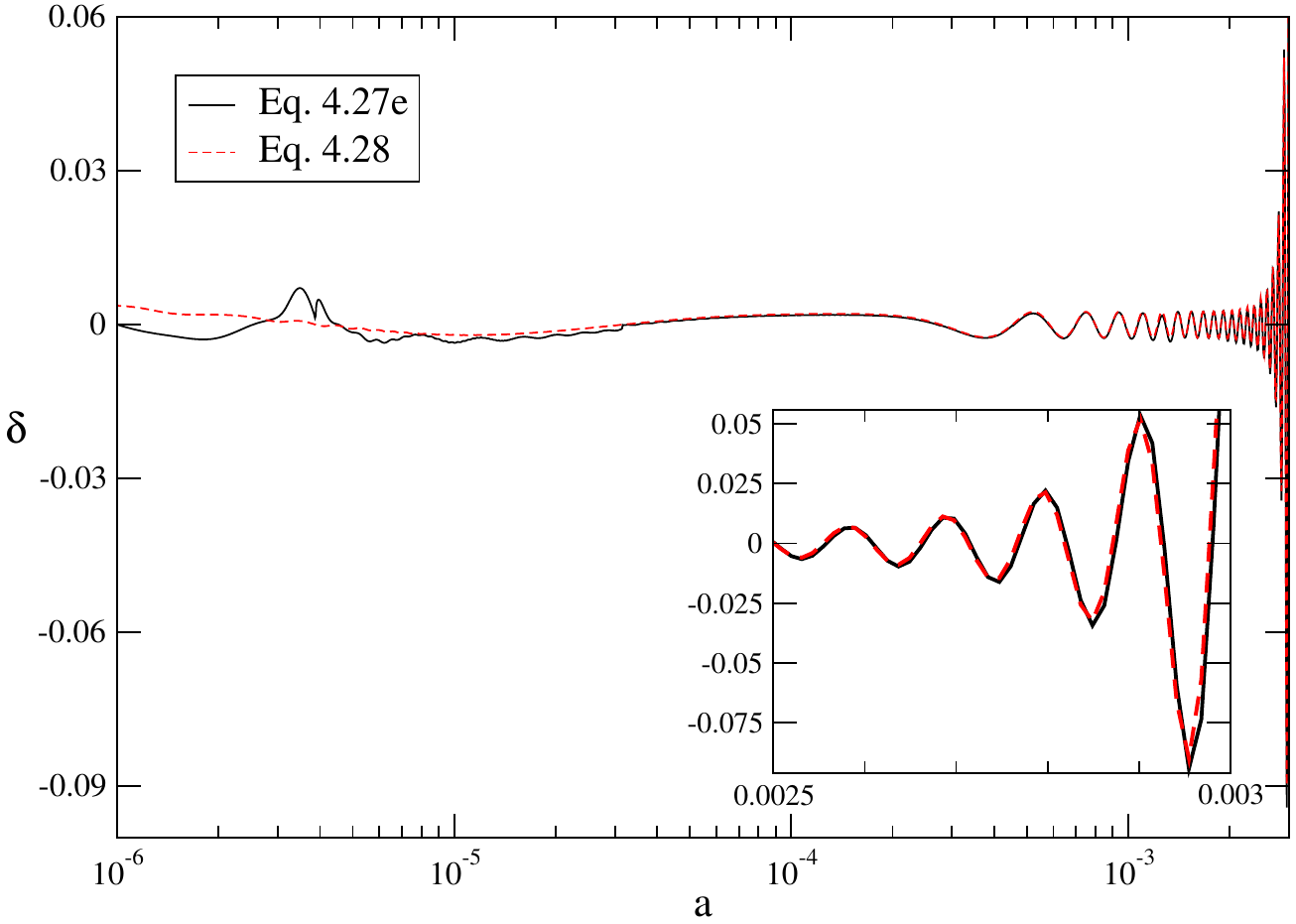}}
 \caption{Evolution of the density contrast $\delta$ for a perturbation with wavelength $\lambda_{k}\sim2$Mpc 
calculated with the diferential equation $\ref{eq:deltanum}$
and the algebraical expression $\ref{eq:deltafor}$. The inset shows the oscillations of the density constrast
when $a>10^{-3}$.}
 \label{fig:deltalin}
\end{figure}

Finally, Fig.~\ref{fig:deltalin} shows the evolution of the density contrast as a function of $a$. It was obtained using the
differential equation (\ref{eq:deltanum}) and the algebraical  
expression (\ref{eq:deltafor}).
We see that the density contrast evolves slowly at early stages 
of the Universe. This Figure clearly shows how it begins to grow before recombination ($a\sim 10^{-3}$). 
These small fluctuations in the SFDM density contrast can be sufficient 
to lead to structure formation in the Universe. Therefore,
overdense regions might be able to form galactic halos at earlier times than those proposed in the standard model. 
During and after the recombination era the density contrast appears to grow in the same way as in the $\Lambda$CDM profile, i.e., 
the SFDM density contrast grows in a very similar fashion
as it does in the $\Lambda$CDM model. 
This can be seen by looking at the temporal average of the terms $F_{\Phi}$ and $G_{\phi}$ in equation (\ref{eq:deltlini}). 
From Fig.~\ref{fig:FPhi0} we see that $\left< F_{\Phi} \right>$ tends to $1$. 
On the other hand, $\left<G_{\phi}\right>$ drops to zero, implying that the second term on the right-hand side of Eq.~(\ref{eq:deltlini}) 
is zero (see Fig.~\ref{fig:GPhi}). Since the time average of the perturbed scalar pressure $\left<\delta P_{\Phi}\right>\rightarrow 0$ 
(see Fig.~\ref{fig:dp}) and that of $\left<\omega_{\Phi_{0}}\right>\rightarrow 0$ (see Fig.~\ref{fig:w0}), 
we find that Eq.~(\ref{eq:deltlini}) resembles the equation for 
the density contrast as in the CDM model, \cite{chung}. 
This means that SFDM perturbations grow exactly as cold dark matter perturbations, 
only when the SF has reached one of its minima and mimics the behavior of the $\Phi^2$ potential.

It is important to point out that the numerical evolution of system 
(\ref{eq:dslin}) is complicated due to the stark oscillations of the scalar field. 
However, in the absence of these oscillations, the evolution of the scalar perturbations grow very similar to those of the standard model. 

\section{The Nonlinear Regime of SF-dark matter perturbations}\label{sec:qlinear}

In this section we study the evolution of the scalar perturbations in the nonlinear (NL) regime when $\delta\gg1$.
A first study on gravitational instability of scalar fields in the linear and nonlinear regimes was developed in \citep{khlopov}.
Recently, \citet{woo} have performed a numerical study of structure formation with SFDM/BEC model. 
Here, we made our analysis within the framework of the spherical collapse model \cite{pad}. 
This formalism is very useful to understand the structure formation process in the Universe in the nonlinear regime. 
Here we focus on the era where the radiation density is equal to the SFDM
density. At this time $T \ll T_{c}$ and, therefore, we can consider that the scalar potential
has reached the $\Phi_{0}^{2}$ profile with the positive mass term with $m\sim10^{-23}$eV.
We study if $\Phi^{2}$-dark matter perturbations are able to form bound structures as in the standard model.

\subsection{The Spherical Collapse Model}

The spherical collapse model proposed by \cite{gunn} is a simple, but fundamental tool for understanding the growth of fluctuations in the Universe. 
This model considers that the formation of gravitational structures in the Universe can be described by the evolution of an overdense spherical region. 
At early times this region evolves together with the background Universe. However, at a certain time it decouples from the general expansion, slows down, 
reaches a maximum radius (this time is called {\it turn-around} (ta)) and eventually collapses, virializes and stabilizes in a finite region. 
There are several studies on spherical collapse with scalar fields in the context of dark energy \cite{wang_stein,weinberg,mota_bruck,horellou,lahav,nunes_mota,wang}.
Nevertheless, the spherical collapse model for scalar field as dark matter has not been studied.

We begin our analysis by considering a spherical overdense region (cluster) in a background Universe, as in section \ref{sec:background}. 
This cluster is envisaged as a small perturbation with a top-hat density profile in the scalar field. 
This means that the density of dark matter inside this region is spatially constant. Following the spherical collapse formalism, 
the equations governing the cluster's evolution with a perturbed scalar field $\Phi_{p}$ are given by the Raychauduri equation and by the Klein-Gordon equation
\begin{eqnarray}
\frac{\ddot{R}}{R}&=&-\frac{\kappa^2}{6}\left(\rho_{\Phi_{p}}+\rho_{z}+\rho_{\nu}+\rho_{b}+\rho_{\Lambda} 
 +\,3P_{\Phi_{p}}+3P_{z}+3P_{\nu} +3P_{\Lambda}\right),\label{eq:ray}\\
\ddot{\Phi}_{p}&+&3\frac{\dot{R}}{R}\dot{\Phi}_{p}+V_{p},_{\Phi_{p}}=0,
\label{eq:kleinc}
\end{eqnarray}
\noindent
being $R$ the radius of the cluster, $V_{p}\equiv V(\Phi_{p})$ is the perturbed scalar potential and $\rho_{b}$, $\rho_{z}$ ($P_{z}$), $\rho_{\nu}$ ($P_{\nu}$), $\rho_{\Lambda}$ ($P_{\Lambda}$) are the background densities (pressures) for baryons, radiation, neutrinos and cosmological constant respectively. We set $\Phi_p=\Phi_{0}+\delta\Phi(R_{i},t)$, where $\Phi_{0}=\Phi_{0}(t)$ is the background scalar field and $\delta{\Phi}$ is a small perturbation. Notice that $\delta{\Phi}$ does not depend on the spatial coordinates. Thus the perturbation is homogeneous inside the cluster (top-hat density profile), and it only depends on time and on the initial radius $R_{i}$ of the cluster. The perturbed scalar density $\rho_{\Phi_{p}}$ and pressure $P_{\Phi_{p}}$ inside the cluster are defined as $\rho_{\Phi_{0}}+ \delta \rho_{\Phi}$ and $P_{\Phi_{0}}+\delta P_{\Phi}$ respectively, where $\delta\rho_{\Phi}$ is given by
\begin{equation}
\delta\rho_{\Phi}=\frac{1}{2}\dot{\delta\Phi}^{2}+\dot{\Phi}_{0}\dot{\delta\Phi}+m^{2}\Phi_{0}\delta\Phi.
\label{eq:delrho}
\end{equation}
In addition, $\delta P_{\Phi}$ is
\begin{equation}
\delta P_{\Phi}=\frac{1}{2}\dot{\delta\Phi}^{2}+\dot{\Phi}_{0}\dot{\delta\Phi}-m^{2}\Phi_{0}\delta\Phi.
\label{eq:delp}
\end{equation}

The radiation fields have not been perturbed in the dynamics of the cluster because these perturbations have no growing modes susceptible 
to affect the structure formation. Likewise, we have not included the baryonic perturbations because we assume that the baryons fall in the 
gravitational potential of collapsed dark matter halos. On the other hand, the perturbations of dark energy are not important on 
scales below $100$ Mpc \cite{caldwell,wang}. Moreover, the effect of any kind of dark energy on virilization of dark matter is still under discussion. 

If the scalar field is perturbed, the system of equations for the cluster is
\begin{eqnarray}
\frac{\ddot{R}}{R}&=&-\frac{\kappa^2}{6}\left(2\dot{\Phi}_{0}^2+4\dot{\Phi}_{0}\dot{\delta\Phi}+2\dot{\delta\Phi}^2-2V-2\delta\Phi V,_{\Phi_{0}}
+\rho_{b}+2\rho_{z}+2\rho_{\nu}-2\rho_{\Lambda}\right),\nonumber\\
\ddot{\delta\Phi}&=&-3\frac{\dot{R}}{R}\dot{\delta\Phi}-3\dot{\Phi}_{0}\left(\frac{\dot{R}}{R}-\frac{\dot{a}}{a}\right)-\delta\Phi V,_{\Phi_{0}\Phi_{0}}.
\label{eq:sysc}
\end{eqnarray}

The system of equations for the overdense region assuming the quadratic scalar potential $V=m^{2}\Phi_{0}^{2}/2$ are given by
\begin{eqnarray}
\frac{\ddot{R}}{R}&=&-\frac{\kappa^2}{6}\left(2\dot{\Phi}_{0}^2+4\dot{\Phi}_{0}\dot{\delta\Phi}+2\dot{\delta\Phi}^2-m^{2}\Phi_{0}^2
-2m^2 \Phi_{0}\delta\Phi+\rho_{b} + 2\rho_{z} + 2\rho_{\nu} - 2\rho_{\Lambda}\right),\nonumber\\
\ddot{\delta\Phi}&=&-3\frac{\dot{R}}{R}\dot{\delta\Phi}-3\dot{\Phi_{0}}\left(\frac{\dot{R}}{R}-\frac{\dot{a}}{a}\right)-m^2 \delta\Phi.
\label{eq:sca}
\end{eqnarray}

We define the nonlinear density contrast of $\Phi^{2}$-dark matter at any time
\begin{equation}
\delta_{nl} \equiv \delta\rho_{\Phi} / \rho_{\Phi_{0}},
\label{eq:deltanl}
\end{equation}
where $\rho_{\Phi_0}$ and $\delta\rho_\Phi$ are given in equations (\ref{rhophi0}) and (\ref{eq:delrho}). The EoS, $\omega_{\Phi_{p}}$, 
for the perturbed scalar field $\Phi_{p}$ is $P_{\Phi_{p}}/ \rho_{\Phi_{p}}$. This EoS varies with time as in the case of $\omega_{\Phi_{0}}$.
Moreover, we can obtain a differential equation for the time 
evolution of the nonlinear density contrast by combining equations
(\ref{eq:backsf0}), (\ref{eq:delrho}), (\ref{eq:sca}), (\ref{eq:tderdelrho}) 
and (\ref{eq:sumdrdpnl}). In fact, 
from the time derivative of equation (\ref{eq:delrho}) and using equation (\ref{eq:backsf0}) we find
\begin{equation}
\dot{\delta\rho_{\Phi}}=\ddot{\delta\Phi}\left( \dot{\Phi_{0}} + \dot{\delta\Phi}\right) + \dot{\Phi_{0}}\left(m^{2}\delta\Phi -3H\dot{\delta\Phi}\right).
\label{eq:tderdelrho}
\end{equation}
On the other hand, we know that
%(\ref{rhophi0}), (\ref{pphi0}), (\ref{eq:delrho}) and (\ref{eq:delp}) 
\begin{eqnarray}
\rho_{\Phi_{0}}+P_{\Phi_{0}} &=& \dot{\Phi_{0}}^{2}, \nonumber\\
\delta\rho_{\Phi}+\delta P_{\Phi} &=&  \dot{\delta\Phi}^{2} + 2\dot{\Phi_{0}}\dot{\delta\Phi}.
\label{eq:sumdrdpnl}
\end{eqnarray}
From eqs. (\ref{eq:sca}) and  (\ref{eq:tderdelrho}) and using eq. (\ref{eq:sumdrdpnl}) we obtain the desired equation as
\begin{equation}
\dot{\delta\rho_{\Phi}}=-3\frac{\dot{R}}{R}\left( \delta\rho_{\Phi}+\delta P_{\Phi}\right) + 
3\left(\rho_{\Phi_{0}}+P_{\Phi_{0}}\right)\left(\frac{\dot{R}}{R}-\frac{\dot{a}}{a}\right)-m^{2}\delta\Phi\dot{\delta\Phi}.
\end{equation}
Thus, from the time derivative of equation (\ref{eq:deltanl}), the differential equation for the nonlinear
density contrast reads as
\begin{equation}
\dot{\delta_{nl}}=-3\left[ \frac{\dot{R}}{R}\left( 1 + \frac{\delta P_{\Phi}}{\delta\rho_{\Phi}}\right) + F_{\Phi}H\right]\delta_{nl}
+ 3F_{\Phi}\left( H - \frac{\dot{R}}{R}\right)-I_{\delta\Phi},
\label{eq:dedeltanl}
\end{equation}
where we define the function $I_{\delta\Phi}\equiv m^{2}\delta\Phi \dot{\delta\Phi} /\rho_{\Phi_{0}}$.
\subsection{Virialization}\label{subsec:virialization}

The cluster decouples slowing down from the general expansion of the background Universe and reaches a point of maximum radius $R_{ta}$. 
After that, the cluster continues its evolution and begins to collapse to a singularity. However, this is not a physical behavior because 
during the collapse our assumption of radial infall fails. In a real world, the cluster virializes and forms a bound structure of finite size.
In order to compute the virial radius $R_{vir}$ of the bound structure of $\Phi^{2}$-dark matter we use the energy conservation and the virial theorem for this region. First, we assume energy conservation between turn-around  and virialization time. In this case, 
the equilibrium conditions read
\begin{eqnarray}
E_{Tot}|_{ta}&=&E_{Tot}|_{vir},\nonumber\\
U|_{ta}&=&(T + U)|_{vir},
\label{eq:encon}
\end{eqnarray}
where $T$ is the kinetic energy and $U$ is the potential energy of this region. At turn-around time $\dot{R}|_{ta}=0$ thus $T_{ta}=0$.

On the other hand, the virial theorem holds for the whole system at the virialization time
\begin{equation}
2T_{vir} + U_{vir} =0.
\end{equation}
It is common to relate the kinetic energy $T$ to the the potential energy $U$ by
\begin{equation}
2T_{vir}-\left(R\frac{\partial{U}}{\partial{R}}\right)_{vir}=0.
\label{eq:virial}
\end{equation}
Combining eqs.~(\ref{eq:encon}) and (\ref{eq:virial}), we can obtain the
virial radius $R_{vir}$ of the gravitational structure. To do so, we need to compute the potential energy 
for the whole system including the interaction potential energy between $\Phi^{2}$-dark matter and each component of the whole system \cite{caimmi}. 
However, we assume that the effects of other components can be neglected because their densities are smaller than the dark matter density. Thus, we compute the self-potential energy for $\Phi{^2}$-dark matter only. The potential energy is then \cite{landau}
\begin{equation}
U_{\Phi_{p}}=2\pi\int_{0}^{R}\rho_{\Phi_p}\phi\,r^{2}dr,
\label{eq:potentialU}
\end{equation}
where $\phi$ is the gravitational potential due to $\Phi^{2}$-dark matter inside the cluster. The gravitational potential $\phi$ is given by
\begin{equation}
\phi(r)=-\frac{\kappa^{2}}{4}(1+3\omega_{\Phi_{p}})\rho_{\Phi_{p}}\left(R^{2}-\frac{r^2}{3}\right).
\label{eq:gravpot}
\end{equation}
Note that we have taken also the pressure contribution from the perturbed scalar field into account \citep{mota_bruck,nunes_mota}, 
since the Poisson equation (with relativistic corrections) reads
\begin{equation}
\nabla^{2}\phi=\frac{\kappa^{2}}{2}\left(\rho_{\Phi_{p}} + P_{\Phi_{p}}\right).
\label{eq:poisson}
\end{equation} 
The substitution of eq.(\ref{eq:gravpot}) into eq.(\ref{eq:potentialU}) leads to
\begin{equation}
 U_{\Phi_{p}}=-\frac{16}{15}\pi^{2}\left(1+3\omega_{\Phi_{p}}\right)G\rho_{\Phi_{p}}^{2}R^{5}.
\label{eq:fpotU}
\end{equation}
Combining eqs.(\ref{eq:fpotU}), (\ref{eq:encon}) and (\ref{eq:virial}) yields
\begin{equation}
\left(1+3\omega_{{\Phi_{p}}_{vir}}\right)\rho_{{\Phi_{p}}_{vir}}\eta^{5}=-\frac{2}{3}\left(1+3\omega_{{\Phi_{p}}_{ta}}\right)
\rho_{{\Phi_{p}}_{ta}},
\label{eq:rev}
\end{equation}
where $\eta\equiv R_{vir}/R_{ta}$ is the fractional radius. Note that the right-hand side of the above equation is a constant that we can 
compute. Thus, equation (\ref{eq:rev}) allows us to find $R$, $\rho_{\Phi_{p}}$ and $\omega_{\Phi{p}}$ at virialization.

%\subsection{The rescaled equation for $\Phi^2$-dark matter perturbations in the non-linear regime}

In what follows we obtain dimensionless evolution equations for a cluster. We define the next dimensionless variables

\begin{eqnarray}
r_{1}&\equiv&mR,\qquad\qquad\;
r_{2}\equiv\dot{R}, \qquad\qquad y_{2}=\delta_{nl}\nonumber\\
z_{3}&\equiv&\frac{\kappa}{\sqrt{6}}\frac{m\delta\Phi}{H},\qquad
z_{4}\equiv\frac{\kappa}{\sqrt{6}}\frac{\dot{\delta\Phi}}{H}.
\label{eq:varc}
\end{eqnarray}
Using the above definitions, the evolution equations (\ref{eq:sca}) can be transformed into an autonomous dynamical system with respect to the e-folding number $N$. 
The system of equations reduces to
\begin{subequations}
\begin{eqnarray}
r_{1}'&=&sr_{2},\\
r_{2}'&=&\frac{r_{1}}{s}\left(-2x^{2}-4xz_{4}-2z_{4}^2 + u^{2} +2uz_{3}+l^{2}-\frac{b^2}{2}-z^{2}-\nu^{2}\right),\\
z_{3}'&=&sz_{4}+\frac{3}{2}\Pi z_{3},\\
z_{4}'&=&-3x\left(\frac{r_{2}}{r_{1}}s-1\right)-sz_{3}-\,3\frac{r_{2}}{r_{1}}sz_{4}+\frac{3}{2}\Pi z_{4},\\
y_{2}'&=&-3\left[\frac{r_{2}}{r_{1}}s\left(1+ \frac{z_{4}^{2} + 2xz_{4}-2u z_{3}}{ z_{4}^{2} + 2xz_{4} + 2u z_{3}}\right)+F_{\Phi}\right]y_{2}
 + 3F_{\Phi}\left(1-\frac{r_{2}}{r_{1}}s\right)-\frac{I_{\delta\Phi}}{H},
\label{eq:dsc}
\end{eqnarray}
\end{subequations}
where the functions $s$, $x$, $u$, $l$, $b$, $z$, $\nu$ and $\Pi$ were determined in section \ref{sec:background} and $I_{\delta\Phi}/H=2sz_{3}z_{4}/\Omega_{\Phi_{0}}$. 
The nonlinear density contrast can be written in terms of the background and cluster dimensionless variables as
\begin{equation}
y_{2}=\frac{z_{4}^{2}+2xz_{4}+2uz_{3}}{\Omega_{\Phi_{0}}}.
\end {equation}

\begin{figure}[ht]
 \centering
 \scalebox{0.6}{\includegraphics{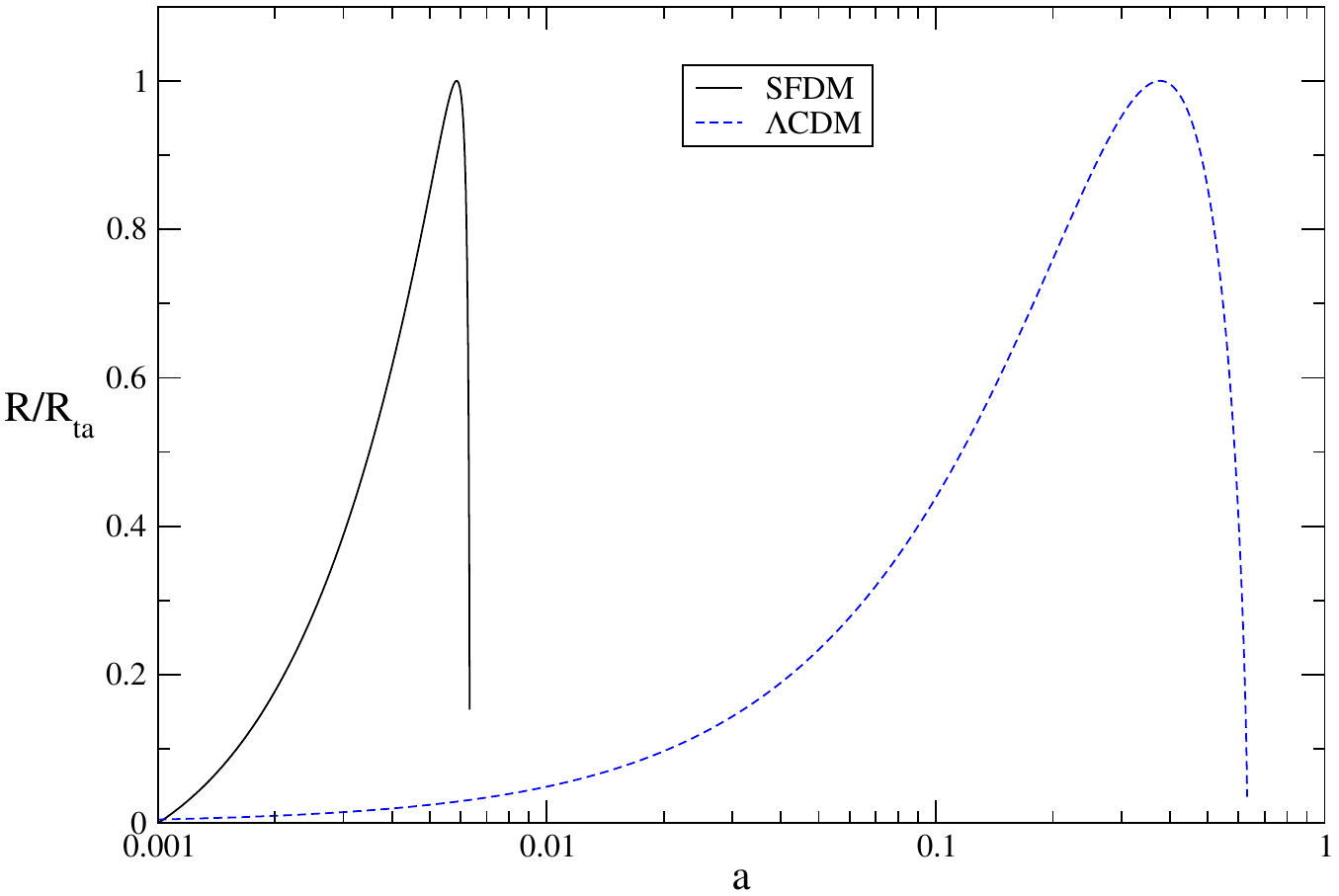}}
 \caption{Evolution of the radius of an initial perturbation 
of $2$ Mpc for SFDM model (black solid line)
and for $\Lambda$CDM model (blue dashed line). 
The radius $R$ has been normalised to the turn-around radius $R_{ta}$. Observe 
how the radius reaches a turn-around point and eventually collapses. 
The singularity is not real because the overdense region virializes
breaking radial symmetry.
The collapse of the fluctuation is earlier in the SFDM model.}
 \label{fig:rcluster}
\end{figure}

\begin{figure}[ht]
 \centering
 \scalebox{0.6}{\includegraphics{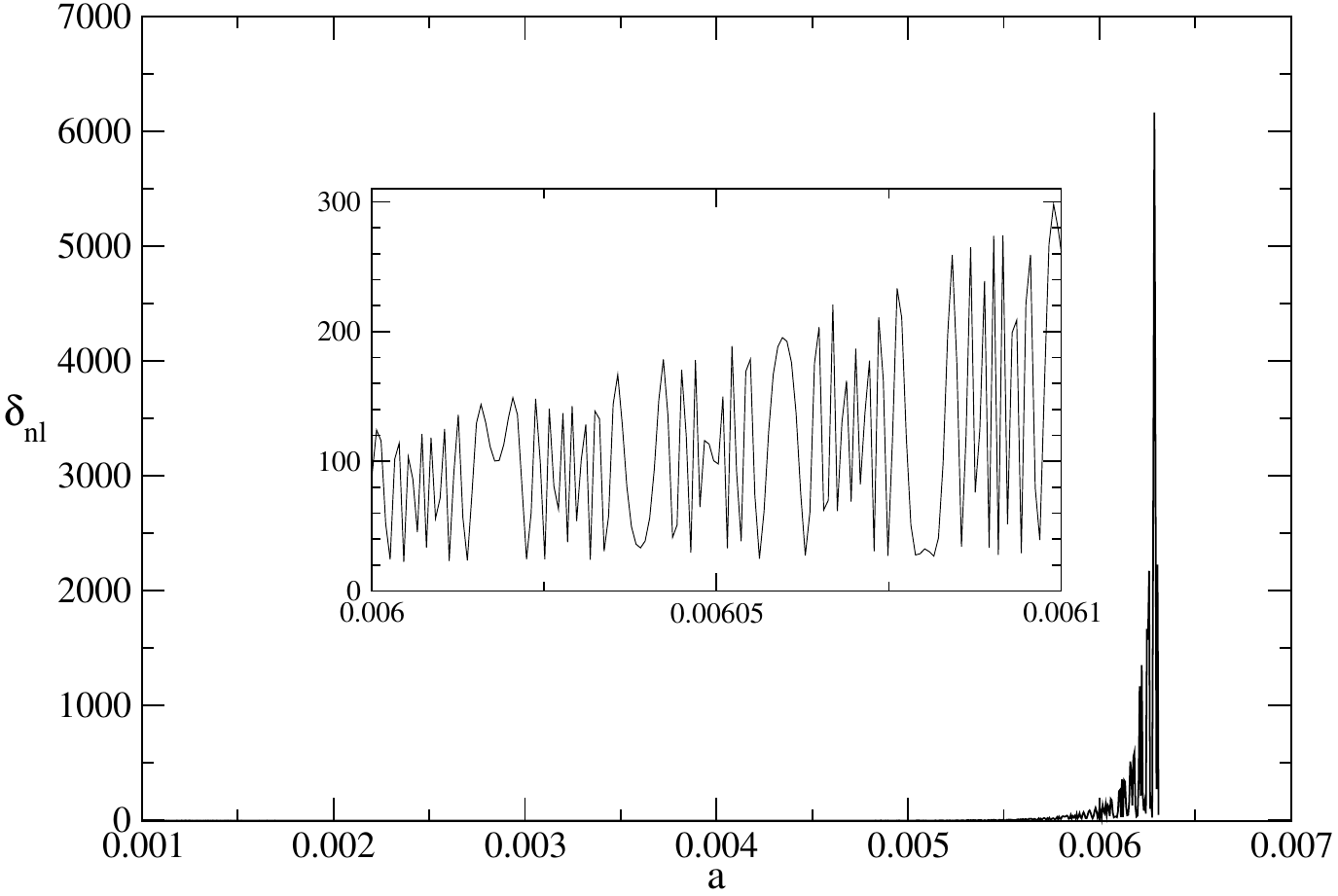}}
 \caption{Evolution of the density contrast in the nonlinear regime
in the $\Phi^{2}$-dark matter model. The inset shows the quick oscillations of the density contrast
close to the virialization time.}
 \label{fig:nldelta}
\end{figure}

In what follows we integrate the dimensionless equations (\ref{eq:dsc}) to investigate if $\Phi^{2}$-dark matter perturbations collapse to form gravitational structures.
%\subsection{Numerical Results for $\Phi^2$-dark matter perturbations in the non-linear regime}
\noindent
Once we have solved the evolution of the background Universe, we can obtain a numerical solution for the system of equations (\ref{eq:dsc}). 
In order to do so, we take the initial conditions in the recombination era at $a_{i}=10^{-3}$. We start with an initial perturbation of $2$ Mpc 
radius and density contrast $\delta=1\times10^{-5}$. Fig.~\ref{fig:rcluster} shows $R$ as a function of $a$. 
Early on, $R$ expands with the background Universe. Later on, $R$ reaches a maximum $R_{ta}$ and starts to collapse. 
For this perturbation, the turn around point occurs at $a_{ta} \sim 5.86\times10^{-3}$, i.e.~at redshift $\sim 170$. Then, $R$ begins to decrease 
until it collapses into a singularity. As discussed in section \ref{subsec:virialization}, 
this singularity is not physical because the dark matter was forced to be spherically distributed and to collapse radially.
Fig.~\ref{fig:nldelta} shows the nonlinear density contrast
for $\Phi^{2}$-dark matter paradigm. 
The transition between the linear regime ($\delta\ll1$) and the nonlinear regime ($\delta\gg1$) occurs when the density contrast is equal to the unity ($\delta$=1). 
For our initial perturbation, the transition occurs at $a_{tr}\sim 5.54\times10^{-3}$. 
The density contrast continues growing up to $\delta_{SFDM_{ta}}\sim 49.94$ at turn around. In an Einstein-de Sitter Universe with $\Omega_{CDM}=1$, 
the density contrast at turn-around is $\sim4.6$. If we compare both paradigms, the perturbations at turn-around are more dense in $\Phi^{2}$-dark matter model 
than they are in the cold dark matter paradigm.
\begin{figure}[ht]
 \centering
 \scalebox{0.6}{\includegraphics{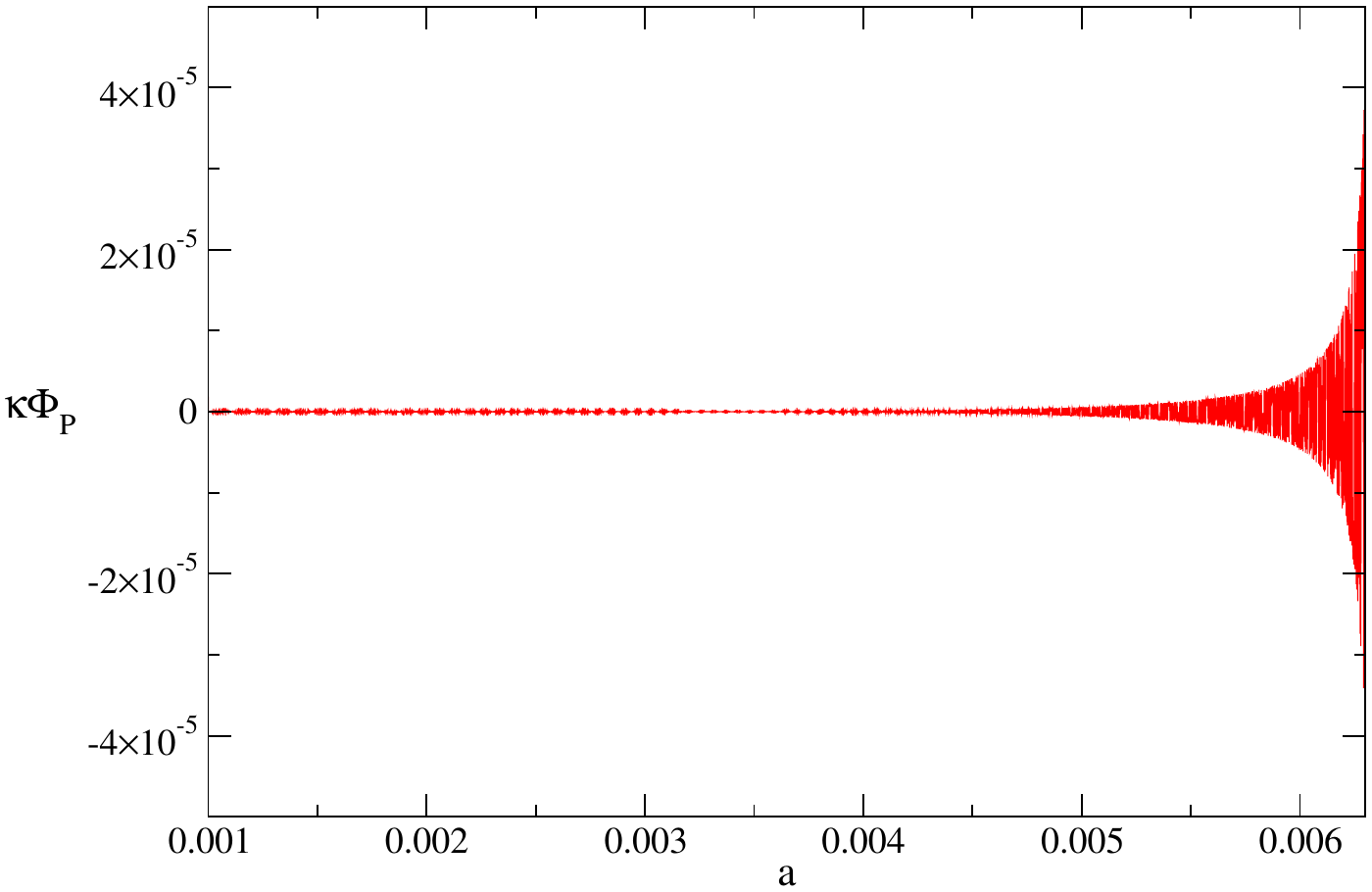}}
 \caption{Evolution of the perturbed scalar field $\Phi_{P}$. Observe that the scalar oscillations grow up in a short time.}
 \label{fig:phiP}
\end{figure}

Now we analyze numerically the relation (\ref{eq:rev}) following the solution until the equality is satisfied. Thus, we
determinate the virialization time and thereby the virial radius and the virial density contrast. We find that the perturbation virializes at 
$ a_{vir}\sim6.3\times10^{-3}$, i.e.~at redshift $z_{vir}\sim158$. This means that the overdensities collapse and virialize earlier in SFDM paradigm 
than in the CDM model and therefore the dark halos are formed at earlier times. 
In order to analyze the cause of this early collapse, we show in Fig.~\ref{fig:phiP} the evolution of the perturbed scalar field $\Phi_{P}$.
Observe that at the recombination era, $\Phi_{P}$ oscillates with very 
tiny amplitudes because the perturbation to the scalar field is very small 
(see Fig.~\ref{fig:delPhi}). However, the amplitude of these oscillations grow in a very short interval
of time ($5\times10^{-3}<a<6\times10^{-3}$). Therefore, the growth of the oscillations in $\Phi_{P}$ is traslated into an 
increase of the perturbed energy density $\rho_{\Phi_{p}}$ and the perturbed scalar pressure $P_{\Phi_{p}}$
in eq.~(\ref{eq:ray}) becomes responsible for the early collapse. 
The reason why the perturbed scalar field inside the overdensity contributes significantly to its collapse,
becomes clear when one analyses the Poisson equation (\ref{eq:poisson}) for the gravitational potential.
It is worth noticing that in the case of CDM perturbations, the pressure is negligible. 
As a result, while the scalar perturbations in the linear
regime behave as those in the standard model, the scalar
pressure, in the nonlinear
regime, plays an important role in the collapse of gravitational structures. 
This is a prediction of scalar field dark matter 
and BEC-type dark matter. Although this result can be controversial, there is some evidence that suggests that dark halos could 
be formed at high redshift \cite{cimatti}. The density contrast at virialization time is between $\delta_{vir}\sim240-3680$. 
These values are larger than the predictions of CDM model. On the other hand, we found that the fractional radius $\eta$ in our model is $\eta=0.43$. 
It is worthwhile noting that this value is below the value in an Einstein-de Sitter Universe where $\eta=0.5$. 
Hence, $\Phi^{2}$-dark matter perturbations can reach equilibrium with a smaller radius.

\section {Conclusions} \label{sec:conclusions}

The SFDM/BEC model could be a serious alternative to the dark matter problem in the Universe. This model predicts that the density profile of galaxies is not cuspy, 
in concordance with observations in LSB and dwarf galaxies density profiles \cite{argelia,victor}. If the existence of cored halos is confirmed, this could be an indication in favour of the SFDM model. In the cosmological regime, the SFDM/BEC and the standard model predict exactly the same physics (see Matos et al. 2009). In this work we have studied the growth and virialization of $\Phi^2$-dark matter perturbations in the linear and nonlinear regimes. Within the linear theory of scalar perturbations we obtain an equation for the evolution of the density contrast. This equation differs from the density contrast equation for CDM. 
However, the extra terms tend to the values of the standard equation. Therefore, as a goal of this work, 
we find that the scalar perturbations in this model grow up exactly as the CDM paradigm. 
Following the spherical collapse model, we also study the nonlinear regime of the evolution of $\Phi^{2}$-dark matter perturbations. 
Here we have shown that the evolution of an overdense region of $\Phi^{2}$-dark matter can collapse and virialize in a bound structure. 
However, we found that the scalar perturbations collapse at earlier times of the Universe and that these virialize with a smaller radius than those in the CDM model.
Thus, the standard and the  SFDM/BEC model can be confronted in their predictions concerning the formation of the first galaxies. 
The existence of massive galaxies at high redshifts is a prediction
of the model and may be used to distinguish
between SFDM/BEC paradigm and CDM. 

\section{Acknowledgments}
The authors wish to thank to V. \'Avila-Reese, X. Hern\'andez
and O. Valenzuela for many helpful discussions, and the referee
for an insightful report that helped improve the clarity
of our paper.
The numerical computations were carried out in the
"Laboratorio de Super-C\'omputo Astrof\'isico (LaSumA) del Cinvestav",
and in the UNAM's cluster Kan-Balam. This work was partially supported
by CONACyT M\'exico under grant no. 166212, I0101/131/07
C-234/07 of the Instituto Avanzado de Cosmologia (IAC) collaboration and 
by the projects CONACyT 165584 and PAPIIT IN106212. 
AS is supported by a CONACyT scholarship.

%\end{quotation}
\end{document}